\documentstyle [12pt,epsf] {article}

\begin{document}
\begin{titlepage}

\begin{center}
\vspace{2cm}
\LARGE
Stellar Masses and Star Formation Histories for $10^5$  Galaxies from the
Sloan Digital Sky Survey
\\                                                     
\vspace{1.5cm} 
\large
Guinevere Kauffmann$^1$, Timothy M. Heckman$^2$, Simon D.M. White$^1$ , 
St\'ephane Charlot$^{1,3}$, Christy Tremonti$^2$,
Jarle Brinchmann$^1$,\\
  Gustavo Bruzual$^4$,
Eric W. Peng$^2$,
Mark Seibert$^2$,\\ 
Mariangela Bernardi$^5$, Michael Blanton$^6$,  Jon Brinkmann$^7$,
Francisco Castander$^8$, 
Istvan Cs\'abai$^{9,2}$, Masataka Fukugita$^{10}$,
Zeljko Ivezic$^{11}$, Jeffrey A. Munn $^{12}$, Robert C. Nichol$^{13}$,
Nikhil Padmanabhan$^{14}$, 
Aniruddha R. Thakar$^2$,
 David H. Weinberg$^{15}$, Donald York$^{5}$
\\
\end{center}
\vspace{5.3cm}
\small
{\em $^1$Max-Planck Institut f\"{u}r Astrophysik, D-85748 Garching, Germany} \\
{\em $^2$Department of Physics and Astronomy, Johns Hopkins University, Baltimore, MD 21218}\\
{\em $^3$ Institut d'Astrophysique du CNRS, 98 bis Boulevard Arago, F-75014 Paris, France} \\
{\em $^4$ (AE) Centro de Investigaciones de Astronomia, Merida, Venezuela}\\
{\em $^5$ Department of Astronomy, University of Chicago, 5640 South Ellis
Ave, Chicago, IL 60637}\\ 
{\em $^6$ Department of Physics, New York University, 4 Washington Place, New York, NY 10003}\\ 
{\em $^7$ Apache Point Observatory, P.O. Box 59, Sunspot, NM 88349} \\
{\em $^8$ Department of Physics, Yale University, P.O. Box 208121, New Haven, CT 06520}\\
{\em $^9$ Department of Physics, Eotvos Lorand University, Pf. 32, H-1518
Budapest, Hungary}\\
{\em $^{10}$ Institute for Cosmic Ray Research, University of Tokyo, Chiba  
277-8582, Japan}\\
{\em $^{11}$ Princeton University Observatory, Peyton Hall, Princeton NJ 08544-1001}\\            
{\em $^{12}$ US Naval Observatory, Flagstaff Station, P.O. Box 1149, Flagstaff, AZ 86002}\\       
{\em $^{13}$ Department of Physics, Carnegie Mellon University, 5000 Forbes Ave, Pittsburgh,
PA 15232}\\
{\em $^{14}$ Department of Physics, Princeton University , Princeton NJ 08544}\\            
{\em $^{15}$ Department of Astronomy, Ohio State University, 140 West 18th Avenue, Columbus,
OH 43210}\\
\pagebreak
\Large           
\vspace{3.3cm}
\begin {abstract}
\normalsize
We develop a new method to constrain the star formation histories, dust 
attenuation and stellar masses of galaxies.
It is based on two stellar absorption line indices, the 4000 \AA \hspace{0.1cm}
break strength and the Balmer absorption line index H$\delta_A$.
Together, these indices allow us to constrain the mean stellar ages  
of galaxies and the fractional stellar mass formed in bursts over the past few Gyr.
A comparison with broad band photometry then yields estimates of dust attenuation
and of stellar mass.
We generate a large library of Monte Carlo realizations 
of different star formation histories, including starbursts of varying strength and
a range of metallicities.
We use this library to generate median likelihood estimates of
burst mass fractions, dust attenuation strengths, stellar masses 
and stellar mass-to-light ratios
for a sample of 122,808  galaxies drawn from the Sloan Digital Sky Survey.
The typical 95\% confidence range in our estimated stellar masses is
$\pm$ 40 \%. We study how the stellar 
mass-to-light ratios of galaxies vary as a function of absolute
magnitude, concentration index and photometric pass-band and how dust
attenuation varies as a function of absolute magnitude and 4000 \AA \hspace{0.1cm}
break strength.
We also calculate how the total stellar mass of the present Universe 
is distributed over galaxies 
as a function of their mass, size, concentration, colour, burst mass fraction 
and surface mass density. We find that most of the stellar mass in the local Universe
  resides in galaxies that have, to within a factor of about 2,
  stellar masses $\sim 5\times 10^{10} M_{\odot}$,
  half-light radii $\sim 3$ kpc, and half-light surface mass densities
  $\sim 10^9 M_{\odot}$kpc$^{-2}$.  The distribution of D$_n$(4000) is strongly
  bimodal, showing a clear division between galaxies dominated by
  old stellar populations and galaxies with more recent star formation.
\end {abstract}
\vspace {0.8 cm}
\end {titlepage}
\normalsize

\section{Introduction}
The masses of galaxies are traditionally estimated by dynamical
methods from the kinematics of their stars and gas.
Ever since extended HI rotation curves were first
measured for spirals, it has been clear that the
implied masses include not only the observed material,
but also substantial amounts of dark matter. Indeed, it
is now believed that most galaxies are surrounded by dark
halos which extend to many times their optical radii and
contain an order of magnitude more mass than the visible components.
The properties of these halos can be inferred directly from 
tracers at large radii such as X-ray emitting atmospheres, systems of
satellite galaxies, or weak gravitational distortion of background
galaxy images. In practice, gaseous atmospheres are visible only 
around some bright elliptical galaxies, while the other two techniques 
are too noisy to apply to individual galaxies; they are used
to obtain average halo properties for stacked samples of similar 
galaxies (Zaritsky et al 1993 ; McKay et al 2002).

In order to understand how galaxies formed, we would like to map the
relationship between the properties of the observed baryonic
components of galaxies and those of their dark halos.                 
This mapping should yield information about
how the baryons cooled, condensed and turned into stars as the
halo-galaxy systems were assembled, and should clarify the complex
physical processes that regulated the efficiency and timing of
galaxy formation.

Most studies of the halo-galaxy relation have used luminosity as a
surrogate for total baryon content, and a kinematic 
measure -- peak rotation velocity, stellar velocity dispersion, X-ray
atmosphere temperature -- as a surrogate for halo mass. Both 
surrogates are far from ideal. Kinematic measures are strongly 
affected by the visible components of the galaxy and so have an
uncertain relation to halo mass; only satellite and weak lensing
studies can reliably estimate total halo masses, albeit as an
average over all galaxies in a chosen class. Galaxy luminosity
may not correlate well with stellar mass (the dominant baryonic
component in all but a subset of the smallest systems). There are
strong dependences on the fraction of young stars in the
galaxy and on its dust content. Thus a technique to correct for 
attenuation by dust and to estimate the mass-to-light ratio
of the underlying stellar population is needed to estimate the 
stellar mass of a galaxy. 

It is well-known that luminosities are less
affected by stellar population variations in the near-infrared than in the optical,
and that extinction
corrections are also smaller at longer wavelengths. 
Recently Verheijen (2001) studied the B, R, I and K'-band
Tully-Fisher relations of a volume-limited sample of 49 spiral galaxies in the Ursa Major    
Cluster. He showed that the 
K'-band relation exhibited the 
tightest correlation with rotation velocity, 
suggesting  that the stellar masses of galaxies are linked closely with                
the masses of the dark matter halos in which they formed. The larger scatter in the
shorter wavelength passbands was attributed to variations in star formation history
and dust extinction between different galaxies in the sample.                            

Although near-infrared luminosities are less dependent on star formation history          
than optical luminosities, they nevertheless exhibit some sensitivity to stellar age.
Bell \& de Jong (2001) estimate that the near-infrared mass-to-light  ratios of local               
spirals can vary by as much as a factor of two and propose a correction
based on the optical colours of the galaxies. Brinchmann \& Ellis (2000)
also used colour-based methods to transform the near-infrared magnitudes of a sample
of intermediate redshift galaxies into stellar masses. In both analyses,
it was assumed that the star formation histories of  
galaxies could be described by simple exponential
laws. If galaxies formed a major ($> 10$\%) fraction of their
stars in recent  bursts, it was shown that this would introduce 
substantial errors into the estimated
mass-to-light ratios.

In this paper, we make use of spectral indicators
as diagnostics of the past star formation histories of galaxies. In particular,
two spectral features, the 4000 \AA \hspace{0.1cm}  break  and the  H$\delta$
absorption line, provide important information about the ages of the stellar populations
in galaxies and are able to distinguish recent star formation histories dominated by bursts 
from those that are more continuous. 
We develop a method based on these two indicators and on broad-band photometry
that allows us to
derive maximum likelihood estimates of the stellar mass of a galaxy, 
the attenuation of its starlight by dust, and the fraction of its stars                   
formed in recent bursts. We apply our method to a sample of $\sim 120,000$ galaxies
drawn from the Sloan Digital Sky Survey and show how the derived stellar mass-to-light
ratios of galaxies in four Sloan bandpasses 
($g$, $r$, $i$ and $z$; Fukugita et al 1996)
depend both on galaxy luminosity and on structural parameters such as concentration index.
We also compute the fraction of the total stellar mass in the Universe in galaxies
of different masses, sizes, colours and concentrations. The dependence of galaxy properties
on stellar mass is addressed in a separate paper (Kauffmann et al 2002; Paper II)

\section {Description of the Observational Sample and Spectral Reductions}

The sample of galaxies analyzed in this paper is drawn from the Sloan Digital Sky
Survey. This survey will obtain $u$, $g$, $r$, $i$ and $z$ photometry of
almost a quarter of the sky and spectra of at least 700,000 objects.
The survey goals and procedures are outlined in York et al. (2000) and a 
description of the data products and pipelines is given in Stoughton et al.(2002).

Our sample of galaxies is  drawn from all available spectroscopic observations 
in the SDSS Data Release One (DR1).
We have included all objects spectroscopically classified as galaxies
with Petrosian $r$ band magnitudes
in the range $14.5 < r < 17.77$ after correction for foreground galactic
extinction using the reddening maps of Schlegel, Finkbeiner \& Davis (1998).
This is equivalent to  the `main' galaxy sample designed for large-scale
structure studies.
Our sample contains a total of 122,808 galaxies.    
More details about the SDSS photometric system may be found in 
Fukugita et al (1996) and Smith et al (2002). Information relating to the SDSS
camera and photometric monitoring system can be found in 
Gunn et al (1998) and Hogg et al (2001).
Details of the spectroscopic target selection
of galaxies are given in Strauss et al (2002).

The spectra were reduced using version v4.9.5 of the Spectro2d pipeline.
The processed spectra are wavelength and flux-calibrated.
The spectral indicators (primarily the 4000 \AA \hspace{0.1cm}
break and the H$\delta_A$ index) and the emission lines
(the equivalent widths and fluxes of  H$\alpha$ and H$\beta$)
that are discussed in this paper are calculated using a special-purpose code
described in detail in Tremonti et al (2002, in preparation).
Here we provide a brief summary of the main features of the code.

Galaxies display a very rich stellar absorption
line spectrum, which encodes much information about their stellar
populations, but complicates measurements of  nebular emission lines. In
most early-type galaxies, the [OII], [OIII], and
H$\alpha$ emission lines are similar in strength to fluctuations produced by the many
superposed stellar absorption features.
Although late-type galaxies tend to have stronger emission lines and
spectra dominated by hotter, more featureless stars, stellar
Balmer absorption can still be substantial (2 - 4 \AA\ at H$\beta$).

We therefore perform a careful subtraction of the stellar absorption line spectrum
before measuring the nebular emission lines. This is accomplished by fitting the
emission line free regions of the spectrum with a model spectrum. The models are based on    
the new population synthesis code of 
Bruzual \& Charlot (2002, in preparation, hereafter BC2002) which
incorporates high resolution (3 \AA\ FWHM) stellar libraries.  (Salient
details of the libraries are discussed further in Section 3).
We have generated a set of 39 template spectra spanning a wide range in age and metallicity.
The template spectra are convolved with a Gaussian  
to match the stellar velocity dispersion of each
galaxy and rebinned to the SDSS dispersion ($\Delta$log$\lambda$= $10^4$). The best fitting  
model spectrum  is constructed from a non-negative linear combination of the
template spectra, with dust attenuation modeled as a  $\lambda^{-0.7}$ power law
of adjustable amplitude (Charlot \& Fall 2000).

In Fig. 1, we plot two representative SDSS galaxy spectra over the
wavelength interval 3700-4400 \AA\  (restframe), which
includes both the 4000 \AA\  break and the H$\delta$
absorption features that are used extensively in this paper. 
The upper panel in Fig. 1 shows the spectrum of a
star-forming galaxy with strong Balmer emission lines, while the lower
panel shows a typical early-type galaxy with strong stellar
absorption features. The red lines in Fig.1 show the best fit model to
the stellar absorption spectrum  over the
wavelength interval from 3600 to 6800 \AA. We find that in general, the
BC2002 models reproduce the main stellar absorption features of
galaxies in the SDSS survey extremely well.

A more detailed analysis of the spectral  fits will be presented
in Tremonti et al (2002). We note that the top panel of Fig. 1
demonstrates that in star forming galaxies, high resolution models are
{\em required} in order to separate the Balmer emission lines from the
underlying stellar absorption features in a reliable way.  Without these
models, the accuracy of the emission line measurements is compromised by
an unknown correction for stellar absorption. Likewise, stellar continuum
indices such as H$\delta_A$ will be contaminated by  emission-lines
unless these can be removed.

All magnitudes quoted in this paper are Petrosian magnitudes.
For the SDSS, the Petrosian radius is defined as the largest radius at which the local $r$-band
surface brightness is at least  one fifth the mean surface brightness interior to that radius.
The Petrosian flux is then the total flux within a circular aperture two
times the Petrosian radius. Details about the Petrosian flux measurements are given
in Lupton et al (2002) and Strauss et al (2002). 
The SDSS Petrosian magnitude detects essentially
all the light from galaxies with exponential profiles and more than 80\%
of the light from galaxies with de Vaucouleurs profiles.

Conversions from apparent magnitude to absolute magnitude depend on cosmology
through the distance modulus DM(z) and on galaxy type through the K-correction
K(z):
\begin{equation} M =m -DM(z)-K(z). \end {equation}
We assume a Friedman-Robertson-Walker cosmology with 
$\Omega=0.3$, $\Lambda=0.7$ and H$_0$= 70 km s $^{-1}$ Mpc$^{-1}$.
We calculate the K-corrections $K(z)$ for each galaxy using the routines
in {\tt kcorrect v1\_11} (Blanton et al. 2002). In order to minimize the errors
in this procedure, we K-correct the magnitudes of all galaxies in our sample to $z=0.1$, which
is close to  the median redshift of the galaxies in our sample. 
Blanton et al. have studied the errors in the K-corrections by
comparing the magnitudes produced by the {\tt kcorrect} routine with broad-band magnitudes     
synthesized from the galaxy spectra themselves. The conclusion is that $r$, $i$ and $z$ band magnitudes
can be reconstructed to the level of the photometry ($\sim$ 1 \%). 
The $g$ and $u$-band magnitudes
have considerably larger errors (5\% and 20\%, respectively). We will not consider 
$u$-band magnitudes in this paper.

\begin{figure}
\centerline{
\epsfxsize=13cm \epsfbox{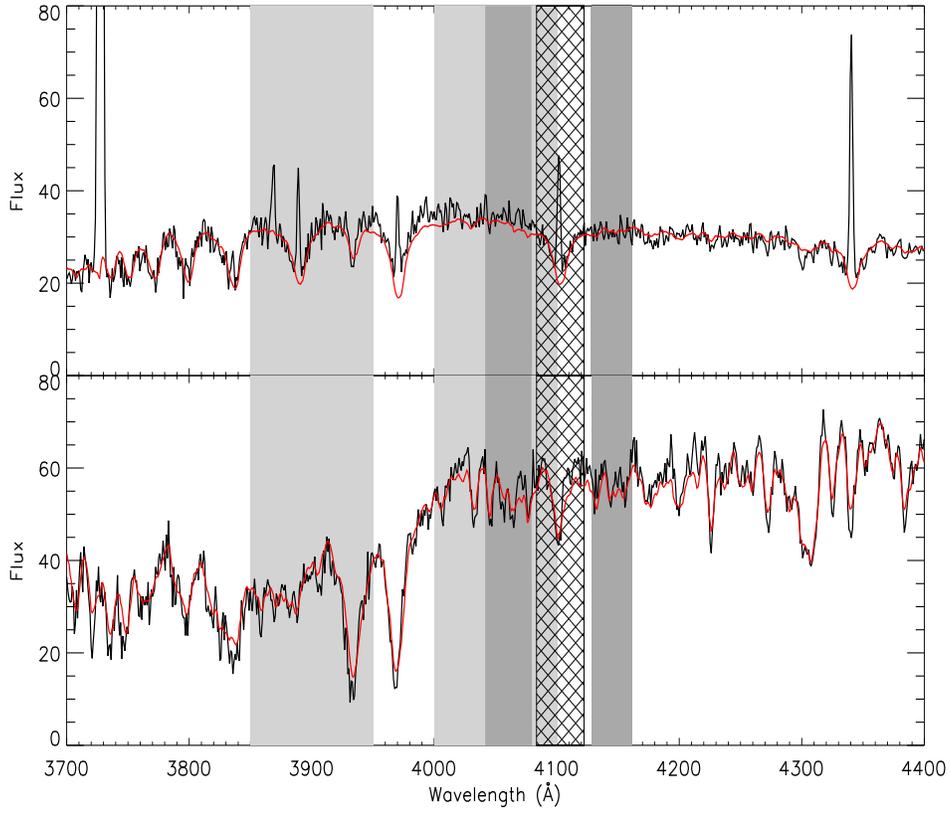}
}
\caption{\label{fig1}
\small
SDSS spectra of a late-type galaxy (top) and an early-type galaxy (bottom) are plotted over the
interval 3700-4400 \AA\ in the restframe. The red line shows our best fit BC2002
model spectrum. The light grey shaded regions indicate the bandpasses over which the
D$_n$(4000) index is measured. The dark grey regions show the pseudocontinua for the
H$\delta_A$ index, while the hatched region shows the H$\delta_A$ bandpass.}
\end {figure}
\normalsize

\section {Spectral Diagnostics of Bursts}

The break occurring at 4000 \AA \hspace{0.1cm} is the 
strongest discontinuity in the optical spectrum
of a galaxy and arises because of the accumulation of a large number of spectral lines
in a narrow wavelength region. The main contribution to the opacity comes from
ionized metals. In hot stars, the elements are multiply ionized and the
opacity decreases, so the 4000 \AA \hspace{0.1cm} break will be small for young stellar
populations and large for old, metal-rich galaxies. A break index D(4000) was defined
by Bruzual (1983) as the ratio of the average flux density F$_\nu$ in the bands
4050-4250 \AA \hspace{0.1cm} and 3750-3950 \AA. 
A definition using  narrower continuum bands 
(3850-3950 \AA \hspace{0.1cm}  and 4000-4100 \AA) was recently introduced by Balogh et al (1999).
The principal advantage of the narrow definition is that the index is considerably
less sensitive to reddening effects. We will adopt the narrow definition as our
standard in this paper and we denote this index as D$_n$(4000).

Strong H$\delta$ absorption lines arise in galaxies that experienced a burst of
star formation that ended $\sim 0.1-1$ Gyr ago. The peak occurs once hot O and B stars,
which have weak intrinsic absorption, have terminated their evolution. The optical
light from the galaxies is then dominated by late-B to early-F stars.
Worthey \& Ottaviani (1997)  defined an H$\delta_A$ index using a central bandpass
bracketed by two pseudo-continuum bandpasses. They parametrized the strength
of this feature as a function of stellar effective temperature, gravity
and metallicity using the Lick/IDS spectral library (Gorgas et al. 1993).
A similar calibration of the 4000 \AA \hspace{0.1cm}
break strength has been carried out by Gorgas et al (1999).                                                     
Most currently available stellar population synthesis models have very low spectral resolution.
The resolution of the  Lick/IDS spectral library is
9 \AA\, a factor of  4 lower than that  of the SDSS spectra. 
The Bruzual \& Charlot (1993) models have a 20 \AA\  spectral resolution,
far too low to measure the H$\delta$ absorption
feature and D$_n$(4000) index reliably. 

High resolution stellar spectral libraries spanning a wide range in metallicity, spectral type
and luminosity class have only recently become available. The STELIB library
(Le Borgne et al 2002) consists of over 250 stars spanning a range of metallicities from
0.05 to 2.5 times solar, a range of spectral types from O5 to M9 and luminosity classes
from I to V. The coverage in spectral type is not uniform at all
metallicities: 
hot ($T_{eff} > 20000$ K) stars of sub-solar metallicity are under-represented, and                 
the library lacks  very cool ($T_{eff} < 3500$ K)  stars of all metallicities. 
The spectra cover the wavelength range from 3500 to 9500 \AA\ 
with a resolution of 3 \AA\  FWHM and a signal-to-noise ratio of $\sim 50$.
Most of the stars in the library were selected  from the catalogue
of Cayrel de Strobel  et al. (1992), which 
contains Fe/H determinations obtained from high resolution 
spectroscopic observations. The selection of stars was optimized to provide good
coverage of the HR diagram. This stellar library has now been incorporated in the
latest version of the Bruzual \& Charlot (1993) population synthesis code.
A full description of the code and of the new stellar library 
will be presented elsewhere (Bruzual \& Charlot 2002).
A description of the underlying stellar evolution prescriptions used in these models
(stellar evolutionary tracks and their semi-empirical extensions) can be found in
Liu, Charlot \& Graham (2000).

The evolution of the D$_n$(4000) and H$\delta_A$ indices 
 following an instantaneous burst of star formation is illustrated
in  Fig. 2. The left panels show the evolution of the two indices for  a solar metallicity single-age stellar population (SSP).
In order to assess how much systematic uncertainty there is in the calibration of these two indices, we have compared
model predictions using 3 different empirical stellar libraries: 1) the STELIB library with 3 \AA\  
resolution (Le Borgne et al 2002),
2) the Pickles (1998) library with 5 \AA\ resolution, and 3) the Jacoby, Hunter \& Christensen (1984)
library with 4.5 \AA\ resolution. Our results show that at fixed metallicity and age,
the  variations in the behaviour of the
indices  due to differences in the input stellar libraries are $\sim 0.05$ for
the D$_n(4000)$ index and $\sim 1$ \AA\  for  H$\delta_A$. For very old stellar populations,
the systematic differences are somewhat larger. In the right panels, we compare the evolution of the
two indices for SSPs of differing metallicity. Results are shown only for
the STELIB library.   There are no other high-resolution libraries
of stars of  non-solar metallicity that span a wide range in wavelength, 
spectral type and luminosity class,
so we are unable to to carry out an analysis of
systematic effects for non-solar models.
As can be seen, the evolution of
D$_n$(4000) does not depend strongly on metallicity until ages of more than $10^9$
years after the burst. The strongest dependence of the H$\delta_A$ index on metallicity also
occurs at old ages.                       
The calibration in Fig.2 does not take into account the effects of the velocity
dispersion of the stars in galaxies. We tested whether this would make any
difference to our index  calibrations and we find that the effects on H$\delta_A$ and D$_n$(4000) 
are negligible (We note that this is {\em not} true for all the  Lick indices).

Although the time evolution of the D$_n$(4000) and H$\delta_A$ indices does 
depend on metallicity,  the {\em locus}
of galaxies in the  D$_n$(4000)/H$\delta_A$ plane is not as sensitive to 
metallicity  and is a powerful diagnostic of whether
galaxies have been forming stars continuously or in bursts over the past 
1-2 Gyr. In addition, these two stellar indices are largely insensitive to
the dust attenuation effects that complicate the interpretation
of broadband colours.  In Fig. 3
we plot H$\delta_A$ as a function of D$_n$(4000) for 
``pure burst'' star formation histories     
and for continuous star formation histories  spanning 
a large range of formation times $t_{form}$  and exponential star 
formation timescales. Results are shown for three different metallicities.
 As can be seen, the continuous models occupy
a narrow band in the D$_n$(4000)/H$\delta_A$ plane. Galaxies with stronger
H$\delta_A$ absorption strength at a given value of D$_n$(4000) must have
formed some fraction of their stars in a recent burst.

\begin{figure}
\centerline{
\epsfxsize=14cm \epsfbox{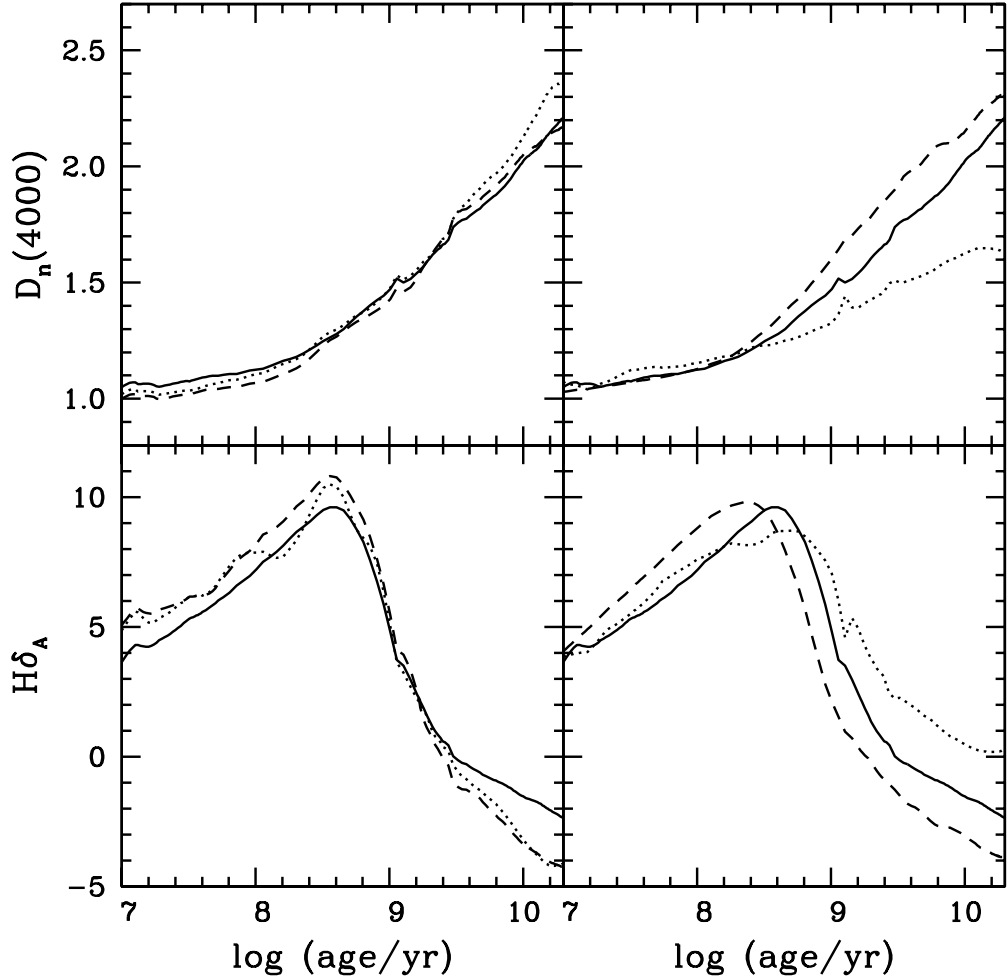}
}
\caption{\label{fig2}
\small
{\bf Left:} The evolution of D$_n$(4000) and H$\delta_A$ following an instantaneous,
solar-metallicity  burst
of star formation. Solid lines show results from BC2002+STELIB, the dotted line
shows results if the Pickles (1998) library is used, and the dashed line is
for the Jacoby, Hunter \& Christensen (1984) library.                                   
{\bf Right:} The evolution of D$_n$(4000) and H$\delta_A$ for bursts of different metallicity.
The solid line is a solar metallicity model, the dotted line is a 20 percent solar model and the
dashed line as a 2.5 solar model.}
\end {figure}
\normalsize

In Fig 3. we also show the region of the 
D$_n$(4000)/H$\delta_A$ plane populated by SDSS galaxies. We show the subset of galaxies
for which the error in the measured value of H$\delta_A$ is less than 0.8 \AA \hspace{0.1cm}
and the error in the D$_n$(4000) index 
is less than $\sim 0.03$. This cut includes $\sim$25,000 galaxies or 20\% of the total sample.
The mean errors on the D$_n$(4000) and H$\delta_A$  indices for the full sample are 0.048 and
1.4 \AA\  respectively.
We have corrected the H$\delta_A$ measurements 
for contamination due to nebular emission.
Because the H$\delta$ emission is so much weaker than  that at H$\alpha$
or H$\beta$, even in the dust-free case, the most robust way to correct
for it is to scale directly from the measured H$\alpha$ flux. 
We use the difference between the observed
H$\alpha$/H$\beta$ emission line fluxes and the dust-free case-B recombination value
(2.86) to calculate  the attenuation of the emission lines. We assume an attenuation law of the
form $\tau_{\lambda} \propto \lambda^{-0.7}$ (Charlot \& Fall 2000). 
It is then straightforward to calculate
the emission correction to the $H\delta_A$ index.
We apply this correction to all galaxies where the H$\alpha$ and $H\beta$ emission lines are 
measured with S/N$>3$. The typical emission correction to H$\delta_A$ for late-type galaxies
with D$_n$(4000)$< 1.4$ is around 1 \AA. For early-type galaxies, it is  smaller.
We have also corrected the D$_n$(4000) index for contamination by the [NeIII] emission line
at 3869 \AA. This only affects a very  small fraction of galaxies -- mainly Seyfert IIs.   
Fig. 3 shows that there is good agreement between the area of the
D$_n$(4000)/H$\delta_A$ plane populated by real galaxies and the values predicted
by the population synthesis models. 
We now investigate the effect of different star formation
histories on the estimated stellar mass-to-light ratios of our galaxies.

\begin{figure}
\centerline{
\epsfxsize=14cm \epsfbox{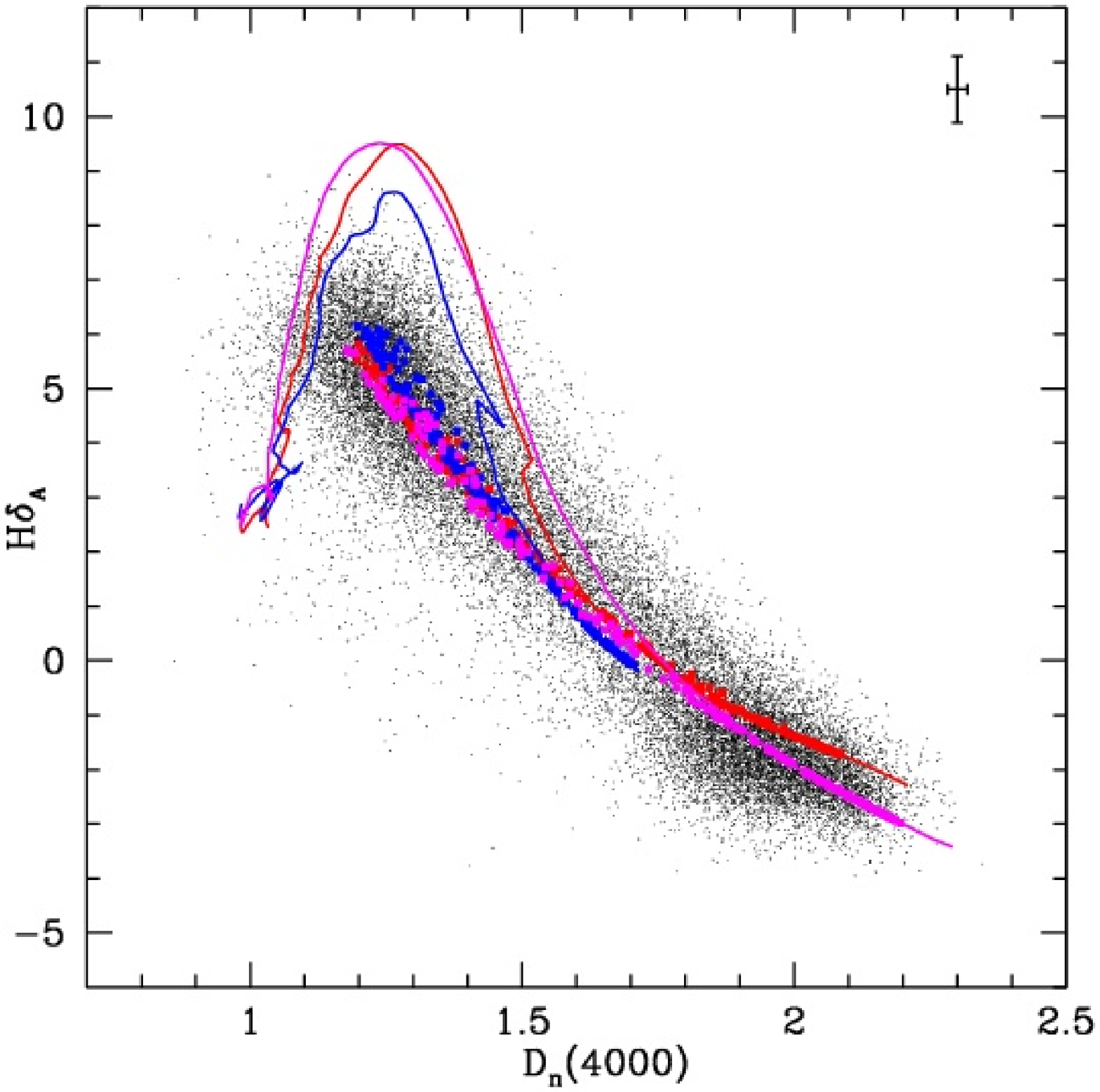}
}
\caption{\label{fig3}
\small
H$\delta_A$ is plotted as a function of D$_n$(4000) for  20\% solar, solar and 2.5 times
solar metallicity bursts (blue, red and magenta lines), and for 20\% solar, solar
and 2.5 solar continuous star formation histories (blue, red and magenta
symbols). A subset of the SDSS data points with small errors are plotted as black dots.
The typical error bar on the observed indices is shown in the top right-hand corner
of the plot. } 
\end {figure}
\normalsize

\section {A Library of Star Formation Histories}

We  use a Bayesian technique to derive estimates of
the  stellar mass-to-light
ratios, dust attenuation corrections and burst mass fractions for each galaxy
in our sample. We  also derive associated confidence intervals for each of
these parameters.
A generalized description of the mathematical basis of our technique is
given in Appendix A.

In Bayesian statistics, an initial assumption is made that the data are randomly drawn 
from a distribution, which is a family of models characterized by a parameter vector {\bf P}.
One has to specify a prior distribution on the space of all possible {\bf P}. This is
a probability density distribution that encodes knowledge about the relative
likelihood of various {\bf P} values in the absence of any data. Typically one takes
a uniform prior in parameters with a small dynamic range and a uniform prior
in the logarithm of parameters with a large dynamic range.

We  set up our prior distribution  by generating a library of Monte Carlo realizations 
of different star formation histories. It is important that our library of models span
the full range of physically plausible star formation histories, and
that it be uniform in the sense that all histories are            
reasonably represented and no a priori implausible corner of parameter space
accounts for a large fraction of the models.

In our library, each star formation history consists of two parts:
\begin {enumerate}
\item An underlying continuous model parametrized by a formation time $t_{form}$ and
and a star formation time scale parameter $\gamma$. Galaxies form stars
according to the law
SFR(t) $ \propto \exp[-\gamma t(\rm{Gyr})]$ from time $t_{form}$
to the present. We take $t_{form}$ to be distributed uniformly over the interval from
the Big Bang to 1.5 Gyr before the present day
and $\gamma$ over the interval 0 to 1. We do not allow star formation rates that increase
with time ($\gamma < 0$) because these are very similar to bursts. 
\item  We  superimpose random bursts on these continuous models. The amplitude of
a burst is parametrized as $A= M_{burst}/M_{cont}$, where $M_{burst}$ is the mass of
stars formed in the burst and $M_{cont}$ is the total mass of stars formed
by the continuous model from time $t_{form}$ to the present.
$A$  is distributed logarithmically from 0.03 to 4.0.  
During the burst, stars form at a constant rate for a
time $t_{burst}$  distributed uniformly in the range  $3\times 10^7$ --
$3 \times 10^{8}$ years. Bursts occur with equal probability at all times   
after $t_{form}$ and we have set the probability  so that 50\%  the galaxies in the library
have experienced a burst in the past 2 Gyr.                
Bursting and continuous models are thus equally represented 
(recall  that   
our stellar indicators H$\delta_A$ and D$_n$(4000) 
are only sensitive to bursts occurring during the past
1-2 Gyr). In section 6.2, we explore to what extent altering the mix of continuous
and bursty star formation histories changes estimated parameters such as the mass-to-light
ratios and burst mass fractions.
\end {enumerate}

We  adopt the universal initial mass function (IMF) as parametrized
by Kroupa (2001). It is very similar in form to the IMF proposed by Kennicutt (1983).
Our models are distributed uniformly in metallicity from
0.25 to 2 times solar. Our final library consists of 32,000 different star
formation histories. For each history, we store a variety of
different parameters, including the predicted values of
D$_n(4000)$, H$\delta_A$, the value of the stellar mass-to-light ratio in the
$z$-band, the $g-r$ and  $r-i$ colours of the stellar populations of the model galaxies, 
and the fraction of
the total stellar mass of the galaxy formed in bursts in the past 2 Gyr
(we will denote this parameter as $F_{burst}$).  
Fig. 4 illustrates how our model galaxies populate the D$_n$(4000)/H$\delta_A$
plane. Even though 50\% of galaxies have had a starburst of varying amplitude in the past 2 Gyr,
most of the models  lie close to the locus of  continuous star
formation histories.
The fact that our models are distributed inhomogeneously in the  
D$_n$(4000)/H$\delta_A$ plane has a significant impact on the confidence intervals we derive
for our parameters. For example, a galaxy will have a relatively         
high probability of being scattered from the continuous to the
bursting region of the plane by observational errors. It will have a much smaller
probability of being scattered the other way around.           
We will illustrate this in more detail in the next section.

\begin{figure}
\centerline{
\epsfxsize=12.5cm \epsfbox{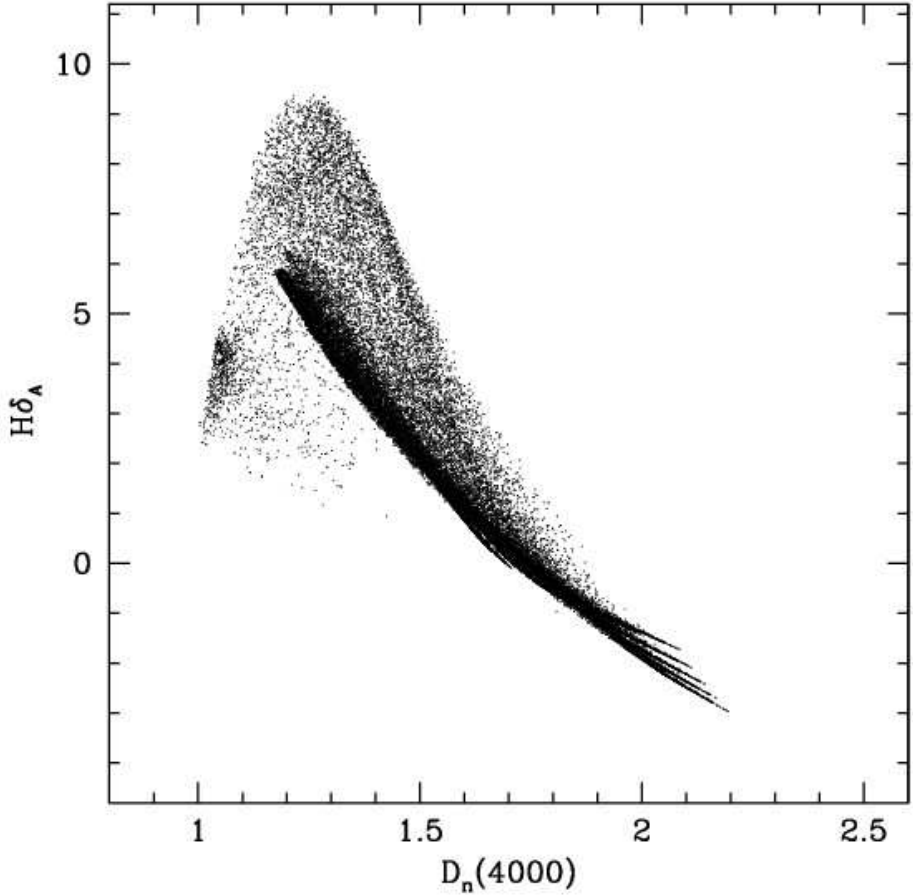}
}
\caption{\label{fig4}
\small
The distribution of galaxies in our model library in the D$_n(4000)$/H$\delta_A$ plane.}
\end {figure}
\normalsize

In the left-hand panel of Fig. 5, we divide the D$_n(4000)$/H$\delta_A$ plane into bins and
we have colour-coded each bin to reflect the average $z$-band mass-to-light
ratio of the simulated galaxies that fall into it. As can be seen, the
mass-to-light ratio is is primarily a sequence in D$_n(4000)$, but at fixed
D$_n(4000)$, galaxies with strong H$\delta$ absorption have a
higher fraction of young stars and hence have smaller mass-to-light ratios. 
As well as average values, one can also compute the
1$\sigma$ dispersion in  these parameters as a function of position in the plane.
This is shown in the right hand panel of 
Fig. 5 and gives an indication of the accuracy with which
mass-to-light ratios could be derived  from the 
models if there were no  
observational errors and in the absence of dust. 
As can be seen, the logarithm of the  $z$-band mass-to-light ratio
can be determined to better than 0.1 dex  for galaxies with D$_n$(4000)$>$ 1.6.
For galaxies with D$_n$(4000)$<$ 1.6, log $(M/L)_{z}$ is still determined
to better than 0.15 dex  for all but the very youngest systems.

In Fig. 6, the bins are colour-coded according to the  {\em fraction} of model SFHs with
$F_{burst}$ in a given range. The blue/cyan areas of the diagram indicate
regions where  more than 95\% of the models in the grid have $F_{burst} > 0.05$.
If a galaxy has measured indices that lie in this region and the errors are small,  
then one can say with high confidence that the galaxy  has experienced a burst in the past 2 Gyr.
In the region  that is colour-coded a  darker shade of  blue,
95\% of galaxies are currently experiencing or have experienced a burst that began no
more than  0.1 Gyr ago. Galaxies in this region 
may be  better termed  ``starburst'' rather than ``post-starburst'' systems.
The green areas of the diagram indicate regions where more than 95\% of the models
have $F_{burst}=0$. If a galaxy lies in  this region,  one can say
with high confidence that a galaxy did {\em not} form a significant fraction of
its stellar population in a burst in the past 2 Gyr. A significant fraction of the
diagram is colour-coded grey. These regions contains a mix of  bursty and
continuous models. The values of $F_{burst}$ derived for galaxies in this region
will depend strongly on how one chooses the mix of star formation histories in the library.
This is discussed in more detail in section 6.2.

For illustrative purposes, we superpose the same subset of SDSS galaxies plotted in Fig.3 on
our colour-coded grid. Once again we see that there is an excellent match between
the area of the diagram occupied by the models and the data. 
There are also a significant number of galaxies that lie within the  blue
bursty region of the plot. A careful analysis of the observational errors is required
before one can assess what fraction of the objects are ``true'' bursty systems.  
This is discussed in more detail in Paper II (Kauffmann et al. 2002).

\begin{figure}
\centerline{
\epsfxsize=17.0cm \epsfbox{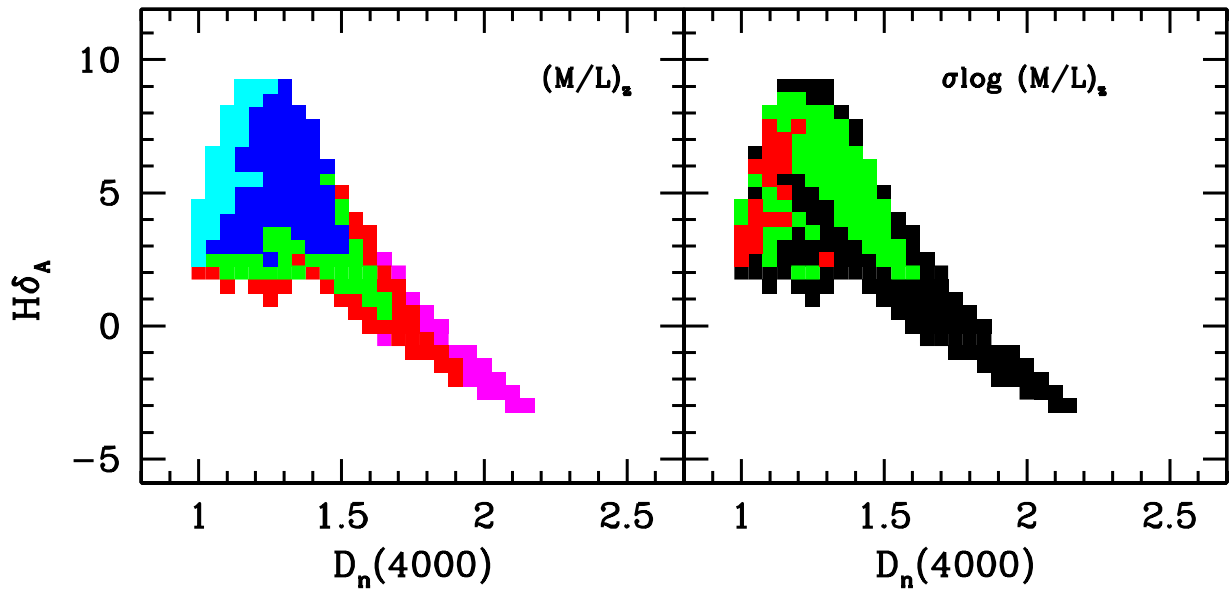}
}
\caption{\label{fig5}
\small
{\em Left:} The D$_n(4000)$/H$\delta_A$ plane has been binned and colour-coded to
reflect the average $z$-band mass-to-light 
ratios 
of the simulated stellar populations that fall into that bin.
Magenta indicates regions with $\log (M/L)_z > 0.2$, 
red is for $0.05<\log (M/L)_{z} < 0.2$,
green is for $-0.1<\log (M/L)_{z}< 0.05$, blue is for $-0.4<\log(M/L)_{z}< -0.1$ and
cyan is for $\log(M/L)_{z} < -0.4$. 
{\em Right:} The plane has been binned and colour-coded to
reflect the 1$\sigma$ dispersion in the  $z$-band mass-to-light 
ratios. 
Black  indicates regions with $\sigma_ {\log (M/L)_z} < 0.1$, 
green is for $0.1< \sigma_ {\log (M/L)_{z}}< 0.15$ and
red is for $\sigma_{ \log(M/L)_{z}} > 0.15$.} 
\end {figure}

\begin{figure}
\centerline{
\epsfxsize=14.0cm \epsfbox{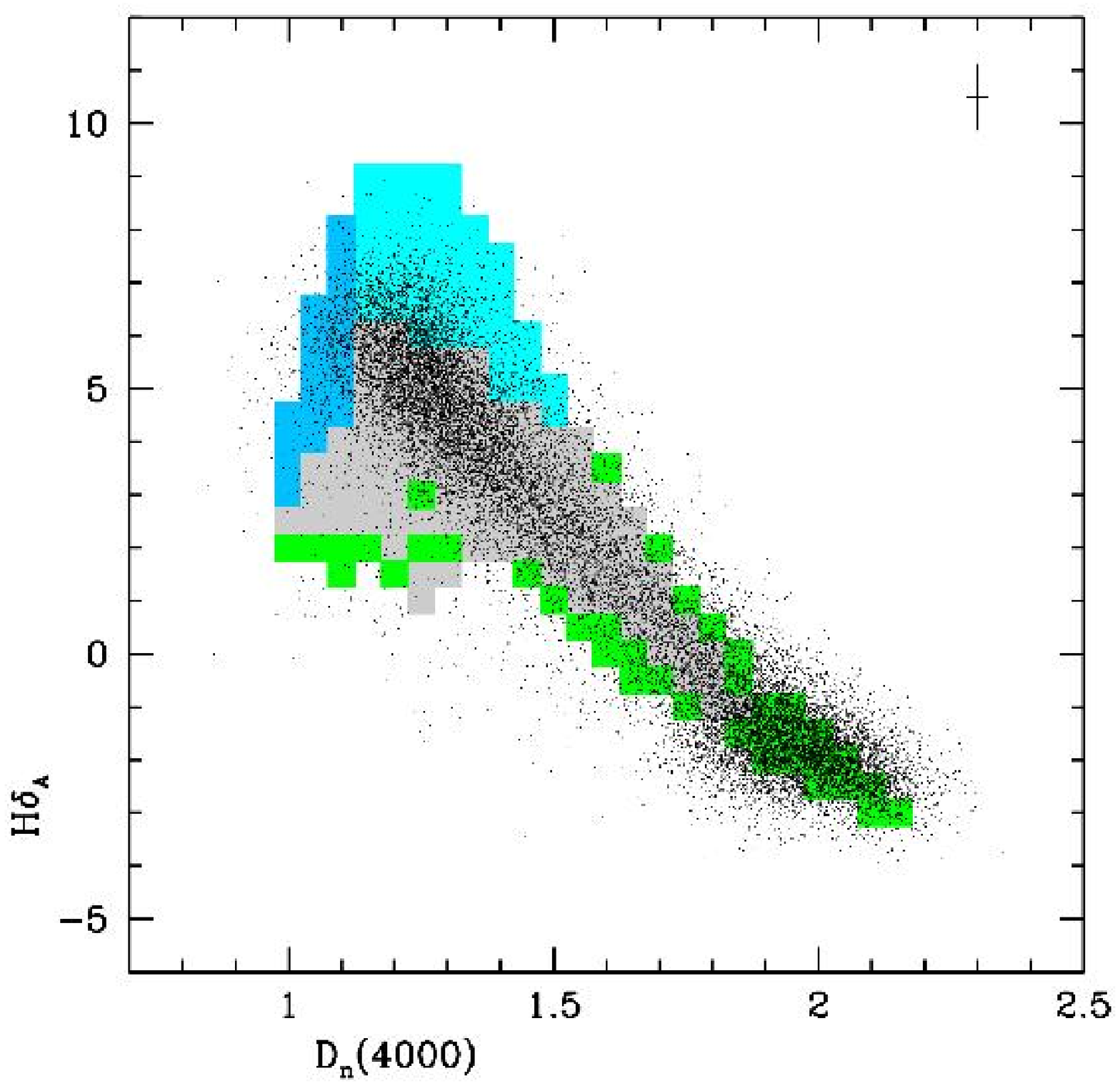}
}
\caption{\label{fig6}
\small
The D$_n(4000)$/H$\delta_A$ plane has been binned and colour-coded to
reflect the fraction of simulated galaxies with $F_{burst}$ in a given range.
Blue indicates regions where 95\% of the model galaxies have $F_{burst}>0.05$ and
the onset of the burst occurred more than 0.1 Gyr ago.  
Cyan indicates regions where 95\% of the model galaxies have $F_{burst}>0.05$ and
the onset of the burst occurred less than 0.1 Gyr ago.  
Green indicates regions where 95\% of the model galaxies have $F_{burst}=0$.
Grey indicates all other regions covered by the model galaxies.}
\end {figure}

\section {Parameter Estimation from the Data}

In this section, we apply our method to our sample of SDSS galaxies
to estimate parameters such as burst mass fractions, stellar masses and dust attenuations.
For each galaxy in the sample, we have a measurement of D$_n$(4000) and H$\delta_A$ as well
as an estimate of the error in these measurements. The likelihood that a galaxy has a given
value  of a parameter  is evaluated by weighting each model in the library
by the probability function $\exp (- \chi^2/2)$ and then binning the probabilities
as a function of the parameter value  (see Appendix A). The most likely value of
the parameter  can be taken
as the peak of this distribution; the most typical value is its median.
We define the 95\% confidence interval by excluding the 2.5\% tails at
each end of the distribution.

\subsection {Estimated versus observed values}

When the observational uncertainty in a spectroscopic index is large,
the best estimate of its true value, taking into account the {\em a priori}
physical constraints in the population synthesis models, may be
significantly different from the observed value.  From now on, we will
use the median value of the derived probability distributions
as our "best" estimate of these parameters (in practice, the peak and
the median usually give similar answers).  It is informative to compare
the {\em estimated} values of D$_n$(4000) and H$\delta_A$ with the {\em observed}
values of the same parameters.  The results of this comparison
are shown in Fig. 7.  For D$_n$(4000), the estimated and observed values
are usually equal to well within the $1\sigma$ observational uncertainty.
 However, the differences between estimated and observed values of
H$\delta_A$ are often considerably larger.  
The reason is that
the models are homogeneously distributed in D$_n(4000)$,
but heterogeneously distributed in H$\delta_A$ (see Fig. 4). As a result, models with  H$\delta_A$
far away from the observed value often contribute significantly to the probability distribution.
This is why the distribution of differences between H$\delta_A$(estimated) and 
H$\delta_A$(observed) shown in Fig. 7 is broad compared to the same distribution derived
for the  D$_n$(4000) index. 

Fig. 8 compares the 2.5 and 97.5 percentile ranges of the estimated                               
values of  D$_n$(4000) and H$\delta_A$ with the observational errorbars. We note that the
observational errors on the D$_n$(4000) index are very small -- typically around 0.02 or 
a few percent of the total range of values spanned by the models. On the other hand,
the H$\delta_A$ index has a typical  error of 1-2 \AA, which is more than 10\%  of the 
range spanned by the models. As can be seen from Fig. 8, the 2.5 and 97.5
percentile ranges of the probability distribution  of D$_n$(4000) correspond extremely well  
to the 2$\sigma$ observational errors. However, the same is not true
for H$\delta_A$. Because the error bars on this index are large, the range of acceptable H$\delta_A$
values is strongly constrained by the region of parameter space occupied by the models.
As a result, 2.5 and 97.5 percentile values
are typically less than 2$\sigma$ away from the best estimate.
We stress that these effects are due to  genuine {\em physical constraints}, 
caused by the fact that H$\delta_A$ can only extend over a limited range of
values for all possible star formation histories.

\begin{figure}
\centerline{
\epsfxsize=12.5cm \epsfbox{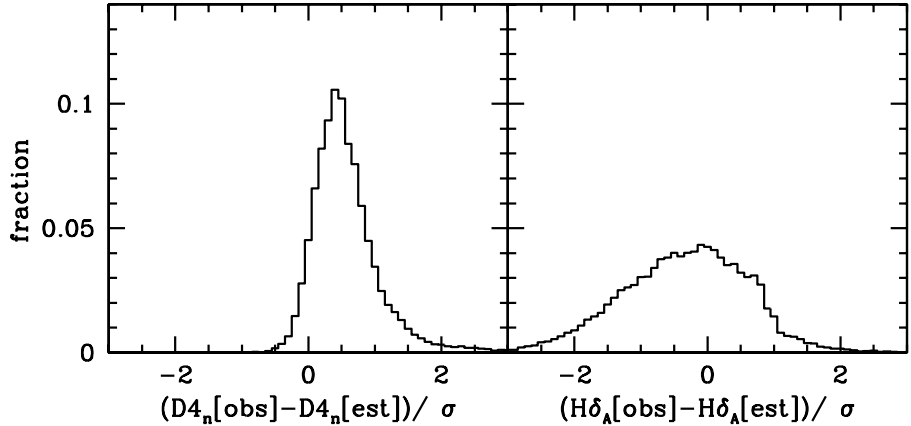}
}
\caption{\label{fig7}
\small
{\em Left:} The distribution of the observed value of D$_n$(4000) minus the estimated one, divided by the
1$\sigma$ measurement error for all galaxies in the SDSS sample.                      
{\em Right:} The same thing except for the H$\delta_A$ index.}                                                
\end {figure}

\begin{figure}
\centerline{
\epsfxsize=13.0cm \epsfbox{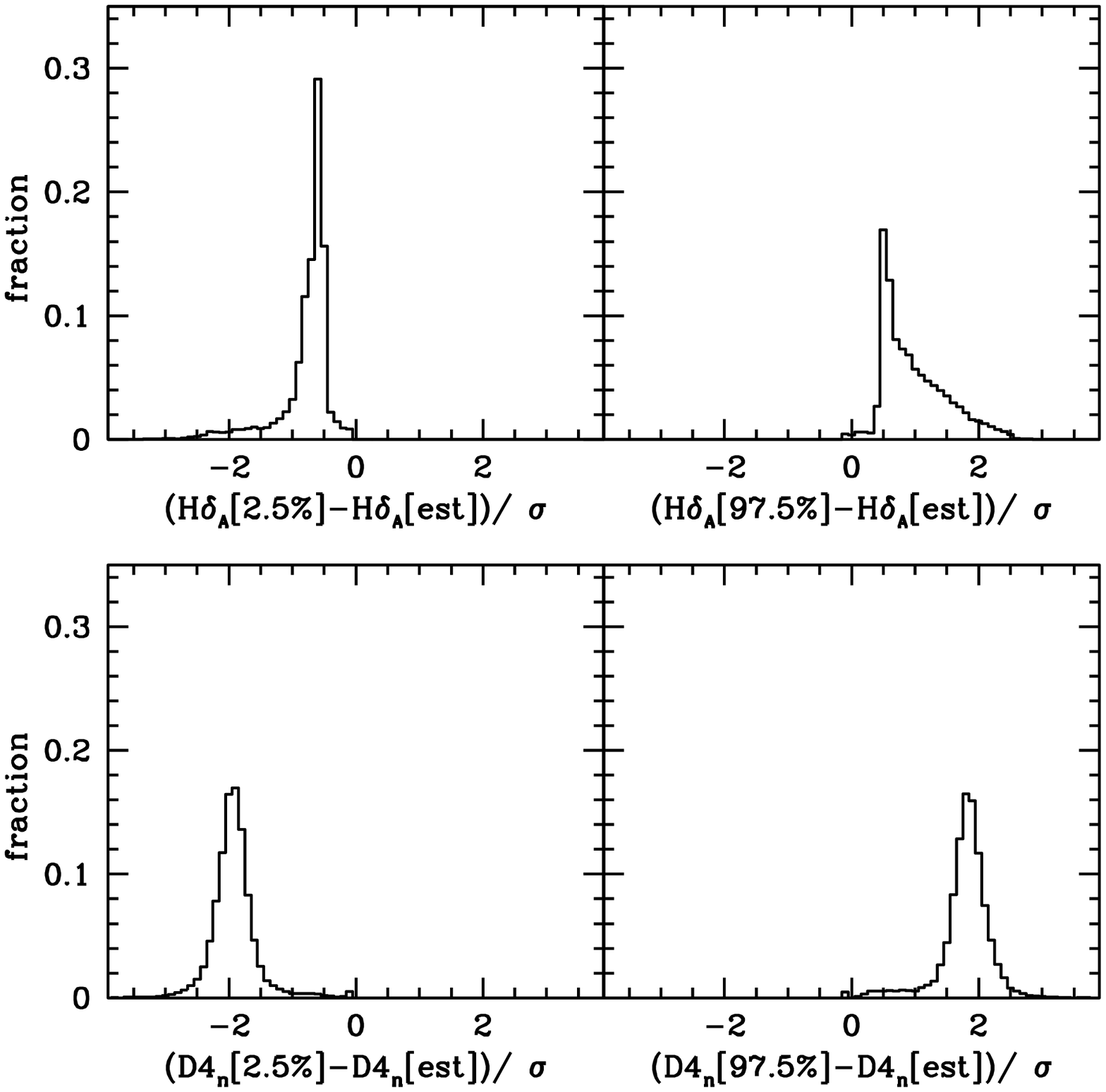}
}
\caption{\label{fig8}
\small
{\em Left:} The 2.5 percentile values of the probability distributions 
of H$\delta_A$ and D$_n$(4000)
minus the best estimates, divided by the 1$\sigma$ measurement error.
{\em Right:} The same  except for the 97.5 percentile value.                                  
These histograms contain the data for our full SDSS sample.}
\end {figure}

\subsection {Estimating Burst Mass Fractions}

Fig. 9 shows the probability distributions of the parameter $F_{burst}$ that we derive for three
galaxies drawn from our observational sample. 

The top galaxy lies in the region of
D$_n$(4000)/H$\delta_A$ parameter space occupied by  galaxies that have formed
stars continuously over the past few Gyr. This is reflected in the derived probability distribution,
which is sharply peaked at a value of $F_{burst}=0$. There is a small tail towards non-zero
values of $F_{burst}$, reflecting the fact that the galaxy  could have had a small
burst of star formation in the past that would not be detectable at the present day. 

The middle galaxy lies well away from the locus of continuous star formation models.
D$_n$(4000) and H$\delta_A$ are both measured with high accuracy. As a result, 
our derived probability distribution indicates that this galaxy {\em must} have formed
more than about 25\% of its stars in a recent burst. The value of $F_{burst}$ is
not very well constrained. This reflects the so-called age/mass degeneracy:
a large burst that occurred long ago is indistinguishable from a smaller burst
that occurred more recently.

The bottom galaxy lies away from the locus of continuous models, but the error on its
H$\delta_A$ index is  much larger than that of the middle galaxy. 
The median value of the probability distribution indicates that a {\em typical} 
galaxy with the same H$\delta_A$ and D$_n$(4000) index measurements  
will have experienced a burst during the past 2 Gyr. However, one
cannot say with any degree of confidence that this particular galaxy must
have undergone such a burst.

Our method thus provides us with an objective way of defining a sample of ``post-starburst''
galaxies from the data. We  study the properties of such a sample in detail in Paper II.

\begin{figure}
\centerline{
\epsfxsize=13.0cm \epsfbox{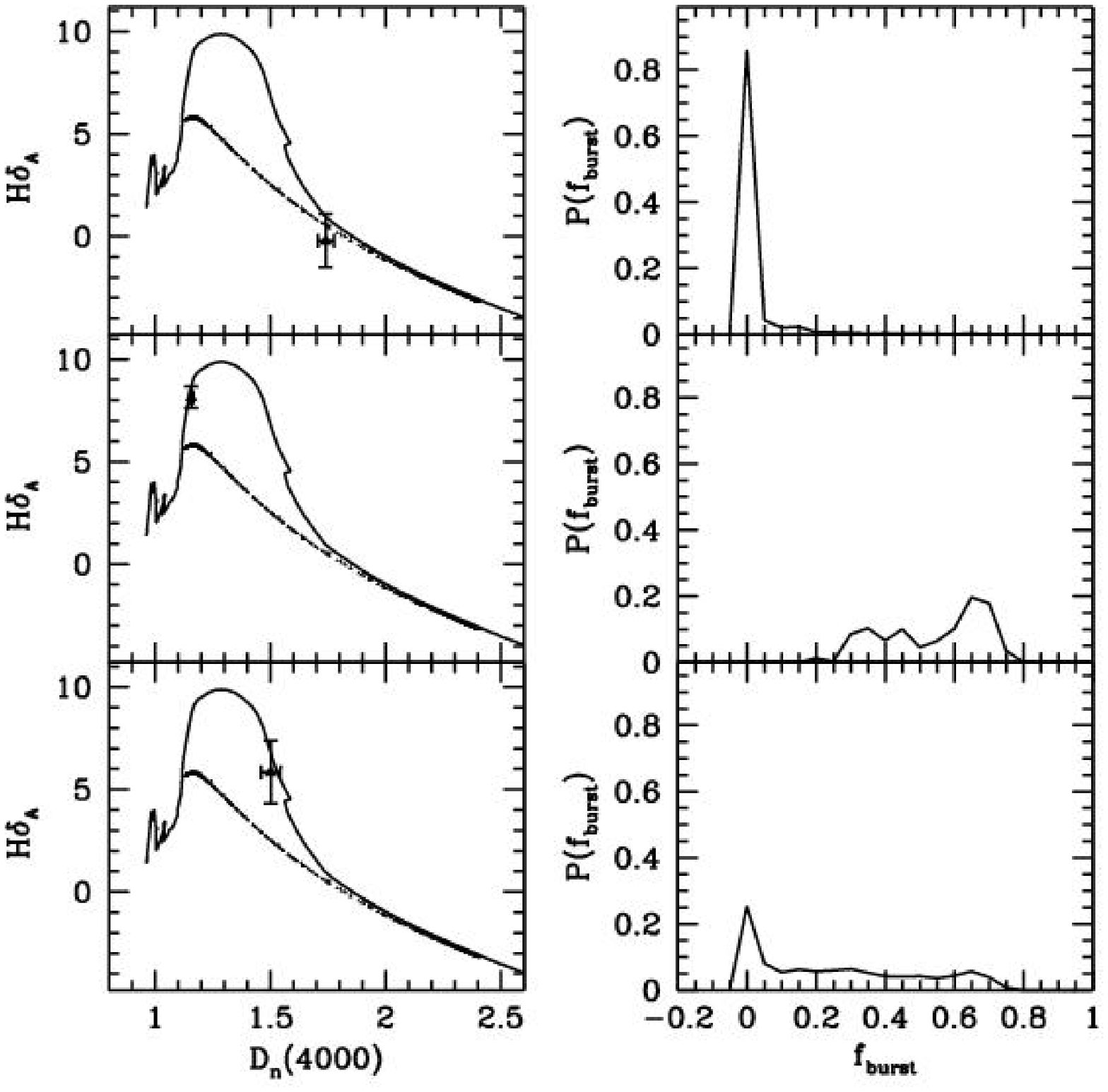}
}
\caption{\label{fig9}
\small
The left-hand panels show the location of three example galaxies in the D$_n$(4000)/H$\delta_A$
plane. Error bars indicate the measurement errors of the D$_n$(4000) and H$\delta_A$
indices. The right-hand panels show the  probability distributions of the
parameter $F_{burst}$ for each galaxy.}
\end {figure}

\subsection {Estimating the Attenuation of Starlight by Dust}

So far, we have only calculated mass-to-light ratios and colours 
of the stellar populations of galaxies.                   
We have not taken into account the attenuation of the starlight
by dust in the interstellar medium. In the $z$-band, dust attentuation
is not expected to be large, but it may nevertheless have substantial
impact on our analysis of  mass-to-light ratios, particularly if certain kinds
of galaxies are much more dusty than others.

We can use our models to estimate the degree of dust attenuation                                
by comparing our best estimate
of the colours of the galaxies from our model library,   
to the observed colour of the galaxy. 
Ideally we should use photometry
that is matched to the fibre aperture to estimate the dust attenuation. 
We have chosen to convolve the spectra
in our sample with the SDSS $g$, $r$ and $i$-band filter functions
in order to generate a set of synthetic ``spectral magnitudes'' that can be compared with
the models. We have tested the accuracy of our spectro-photometry 
by comparing the spectral magnitudes of stars drawn from the SDSS
to the PSF magnitudes output by the photometric pipeline PHOTO (see Tremonti et al
2002 for more details).  We have also subtracted off the main emission lines, including           
H$\alpha$ , [NII] and [SII], to 
generate a set of true ``stellar magnitudes'' for our galaxies.

In Fig.10 we compare the  $g-r$ versus $r-i$
colours generated from the spectra  (black points) with the colours
of the galaxies in our library (blue points). Results are shown for a random subset of
our sample of 120,000 galaxies.
We plot observer-frame colours at two different redshifts: $z=0.03$ and $z=0.11$. In the left panels,
we show colours without any correction for emission lines. The right
panels show what happens when emission lines are removed from the spectra.
Over certain redshift ranges, the effect of emission lines on the colours can be large.           
The $r-i$ colour is strongly affected
by emission at $z \sim 0.1$, because a set of three strong lines (H$\alpha$, [NII] and [SII])  are positioned near the
edges of the $r$ and $i$ passbands.  

The right-hand panels of Fig.10 show that the data is still offset from the models,
even after we have corrected for emission lines. This offset is caused by dust
reddening. The red arrow on the plot indicates the direction of the reddening vector
assuming an attenuation law of the form $\tau_{\lambda} \propto \lambda^{-0.7}$.
The length of the arrow is appropriate for a galaxy with $\tau_V=0.5$, which is
typical for spiral galaxies.
We see that reddening will move galaxies systematically to the right of
the locus occupied by our models. 

We have used the difference between  model colours and the measured colours 
to calculate the reddening for each galaxy in our sample.
By extrapolating to the $z$-band using a standard attenuation curve
of the form $\tau_{\lambda} \propto \lambda ^{-0.7}$, 
we obtain an estimate of $A_{z}$,
the attenuation in the $z$-band expressed in magnitudes. 
We note that the shape of the attenuation curve determined from direct
observations resembles a power law with slope $\sim -0.7$ over a wavelength range from
1250-8000 \AA \hspace{0.1cm} (Calzetti, Kinney \& Storchi-Bergmann 1994). Models based
on a 2-component ISM suggest that the extinction curve may be slightly steeper
than this at longer wavelengths (Charlot \& Fall 2000), but we will use a single
power-law for simplicity. In practice, we find that comparing  $g-r$ or $r-i$              
colours yields very similar attenuation values for most galaxies if the colours  
are corrected for nebular emission. 

The distribution of $A_{z}$ for our full sample of SDSS galaxies is shown in Fig. 11.
The errors in our synthesized magnitudes are extremely small
and the formal 1$\sigma$ error on our estimate
of $A_{z}$ is typically around $\pm$0.12 mag. This does not take into account
systematic errors that may arise as a result of calibration problems. 
Fig. 11 shows that the typical attenuation at $z$-band is quite small.  The median value
of $A_{z}$ is 0.3 mag. Nevertheless there is a long tail to high values of $A_{z}$. More
than 5\% of galaxies in the sample are attenuated by more than a magnitude in the $z$-band.
The dependences of $A_{z}$ on  absolute magnitude  and on D$_n$(4000)
are shown in Fig. 12.  At luminosities below $L_*$, the median attenuation increases
for more luminous galaxies.
At luminosities above $L_*$, the median attenuation drops sharply.  
This can be explained by the fact that the most luminous galaxies are
predominantly early-type systems that contain little dust.
This is shown more clearly in the right hand panel of Fig. 12, where we see that
galaxies with D$_n$(4000)$>$ 1.8 have $A_z=0$ on average.                        
Fig. 12  also shows that the galaxies with the youngest stellar populations 
(small values of D$_n$(4000)) are the most attenuated by dust. This is expected,
because galaxies with  young stars contain more gas and hence more dust than
galaxies with old stellar populations. 

\begin{figure}
\centerline{
\epsfxsize=13.0cm \epsfbox{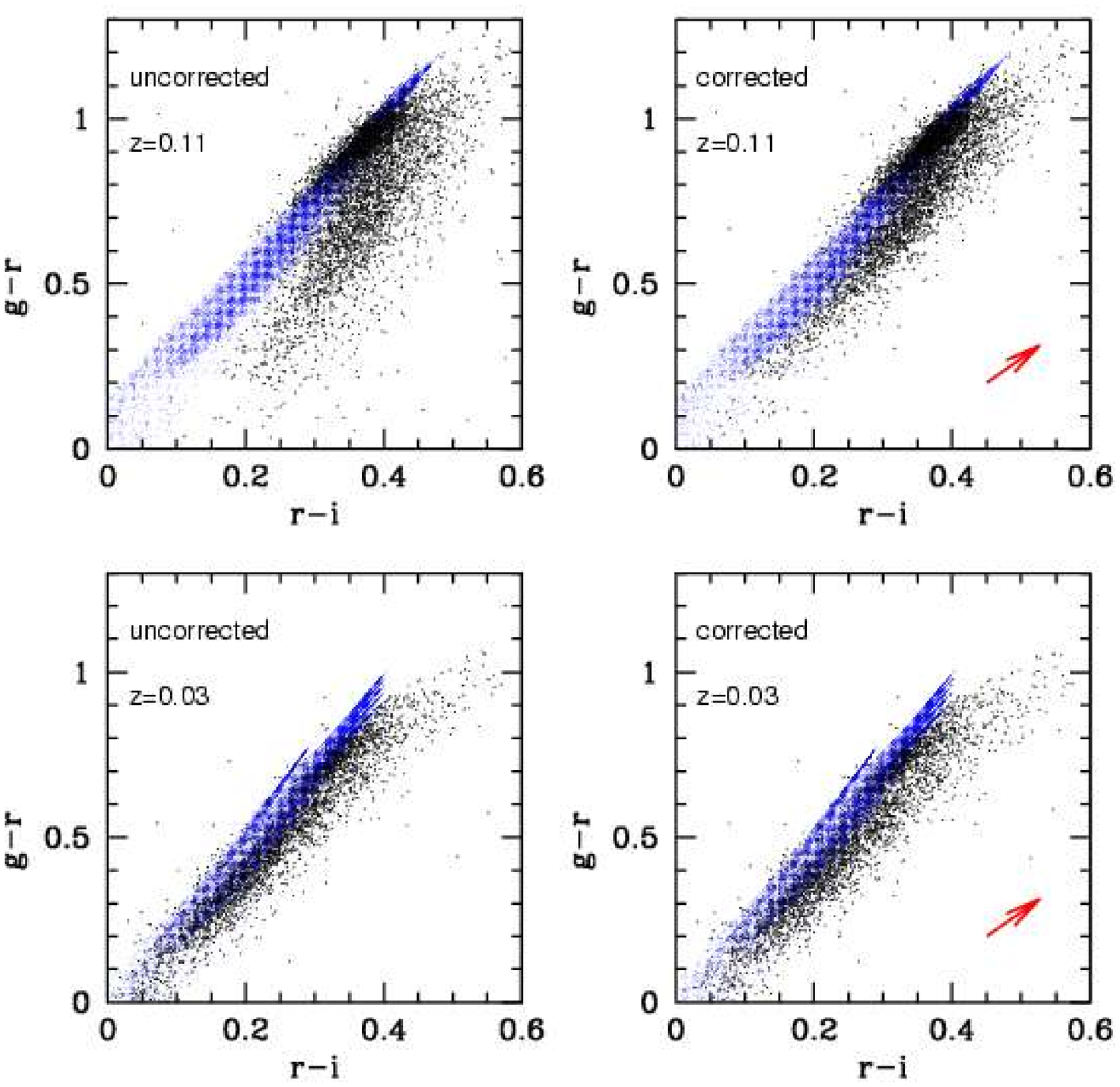}
}
\caption{\label{fig10}
\small
The observed $g-r$ versus $r-i$ spectral colours of a representative subset of
SDSS galaxies (black points) are compared with
our Bruzual-Charlot model grid (blue points) at $z=0.03$ and $z=0.11$. 
The colours have been computed by convolving the spectra with the SDSS filter
functions. In the right panels, the colours are computed after emission lines
have been removed from the spectra. 
The predicted reddening vector assuming an attenuation law of the form 
$\tau_{\lambda} \propto \lambda ^{-0.7}$ is shown as a red arrow.}
\end {figure}

\begin{figure}
\centerline{
\epsfxsize=7.0cm \epsfbox{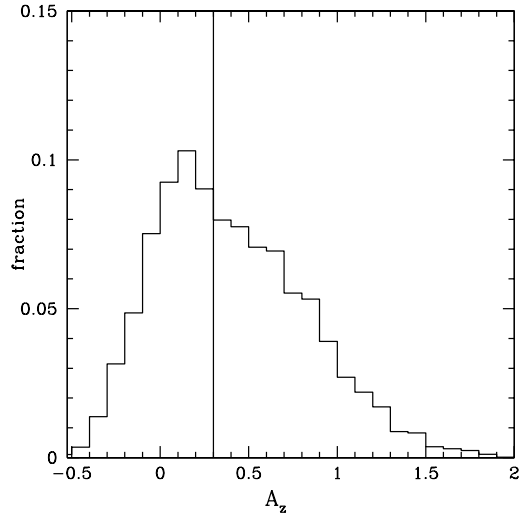}
}
\caption{\label{fig11}
\small
The distribution of the estimated values of the dust attenuation in the  $z$-band                          
for all the galaxies in the sample. The median value is shown as a vertical line.}
\end {figure}

\begin{figure}
\centerline{
\epsfxsize=13.0cm \epsfbox{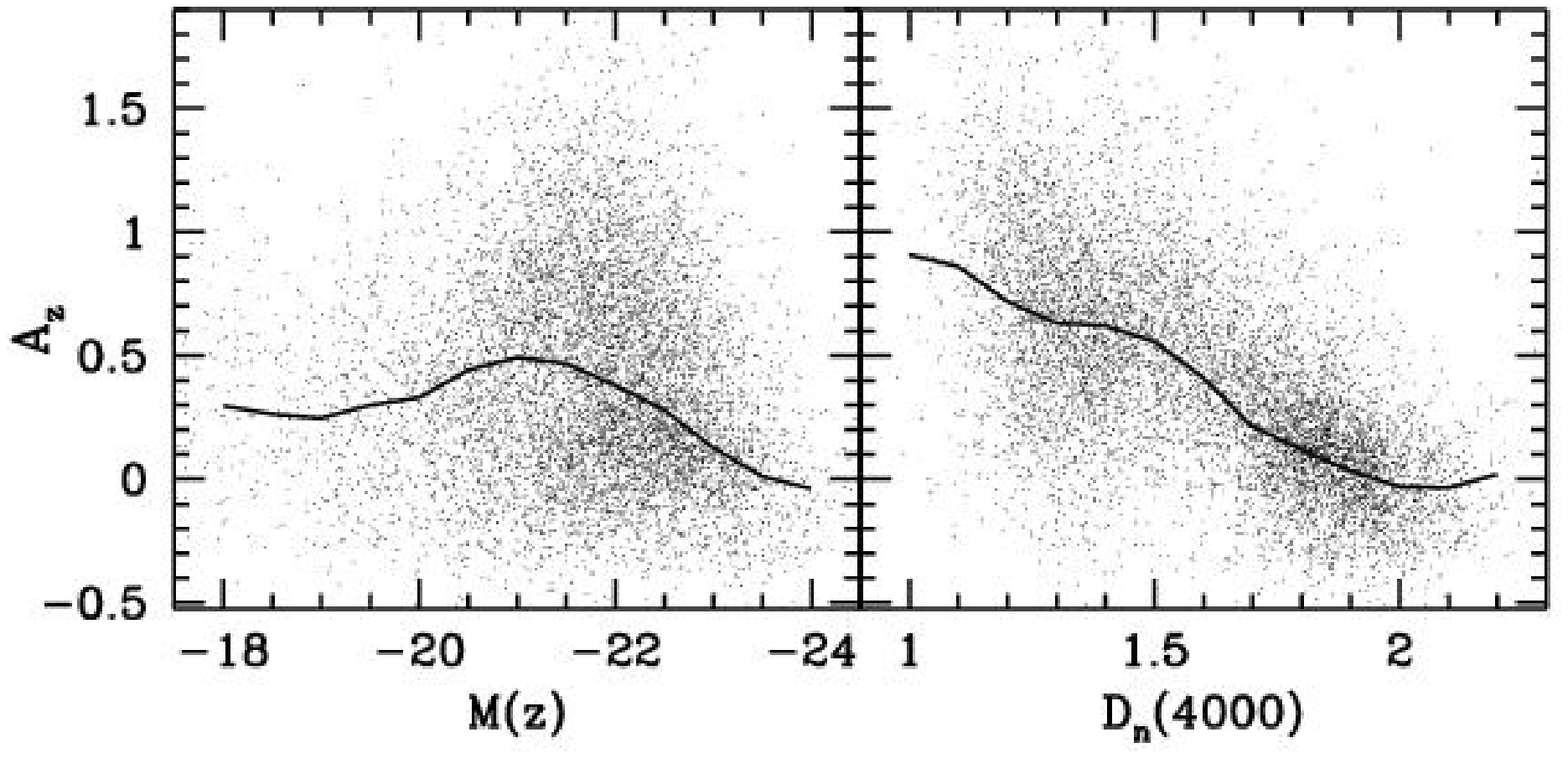}
}
\caption{\label{fig12}
\small
{\em Left:} $A_{z}$ is plotted as a function of $z$-band absolute magnitude                           
for a random subsample of  galaxies . The solid line shows the running median of    
the distribution for the full sample.
{\em Right:} $A_{z}$ is plotted as a function of D$_n$(4000).}                                           
\end {figure}

\subsection {Estimating Stellar Masses and Mass-to-Light Ratios}

In this section, we present estimates of the stellar masses and  
mass-to-light ratios of the galaxies in our sample. To estimate the stellar mass,
we first need to estimate the dust-correction to the observed $z$-band magnitude of the galaxy.
When estimating stellar masses and mass-to-light ratios, 
we  exclude models with $r-i$ colours redder than the
observed colour, because this would give negative attenuation corrections,
which are unphysical.
For each acceptable model, the stellar mass is computed by multiplying 
the dust-corrected luminosity of the
galaxy by the stellar mass-to-light ratio predicted by the model. 
The best estimate of the mass is then obtained by weighting each model by 
the probability function in the usual way. Note that in the calculation of our stellar
masses, we are extrapolating the mass-to-light ratios and $A_z$ values estimated within
the fibre to the galaxy as a whole.

In the top panel of Fig. 13, we plot the $z$-band stellar mass-to-light ratios estimated from the model
as a function of $z$-band absolute magnitude (corrected to $z=0.1$) for all the galaxies in our sample. 
The mass-to-light ratios are plotted in solar units, where we have adopted
$M_{z \odot}(z=0.1)=4.51$ (Blanton et al. 2002).   
As can be seen, the distribution of $(M/L)_z(\rm{model})$ is strongly dependent on galaxy luminosity.
Almost all very luminous galaxies have high mass-to-light ratios. 
Faint galaxies span a  wider range in $(M/L)_z(\rm{model})$. For guidance, we have
marked the value of $(M/L)_z(\rm{model})$ for a galaxy that has formed stars at a constant rate over
a Hubble time as a line on the plot. Galaxies with mass-to-light ratios lower
than this value have had star formation histories that are more weighted towards the
present than continuous star formation lasting  10-12 Gyr. 
Only a few of the most luminous galaxies fall into this category, but at faint
magnitudes many galaxies have formed a large fraction of their stars at late times.

The bottom panel of Fig. 13 shows $(M/L)_z (\rm{model})$ as a function of the standard concentration
parameter $C$, defined as the ratio $R90/R50$, where R90 and R50 are the radii
enclosing 90\% and 50\% of the Petrosian $r$-band luminosity of the galaxy. 
It has been shown
by Shimasaku et al (2001) and Strateva et al (2001) that for bright galaxies, there
is a  good correspondence between concentration parameter and `by-eye' classification
into Hubble type, but there is some disgreement about the value of $C$ that marks the
boundary between early and late type galaxies. Strateva et al (2001) claim that
galaxies with $C> 2.6$ are mostly  early-type systems, whereas
spirals and irregulars have $2 ~< C < 2.6$.  
Not surprisingly, our stellar mass-to-light ratios 
are also  correlated with concentration parameter, with concentrated (early-type)
galaxies exhibiting higher and more uniform values of $(M/L)_z(\rm{model})$ than less concentrated
(late-type) galaxies. 

In Fig. 14, we plot mass-to-light ratios  defined in a different way:
\begin{equation} \frac {M}{L} (\rm{galaxy}) = \frac {Total~ Mass~ in~ Stars} {K-corrected~ luminosity}. \end {equation}
The total mass in stars is computed by multiplying the  dust-corrected and K-corrected
$z$-band luminosity of the galaxy by our estimate of the $z$-band mass-to-light ratio of its
stellar population.  $M/L(\rm{galaxy})$ thus tells us 
how a given  {\em observed} luminosity  translates into stellar mass.
In Fig. 14, we show results in
 four of the five SDSS bands ($g$, $r$, $i$
and $z$). We do not show results for the $u$-band because of difficulties with the K-corrections.
The median value of the mass-to-light ratio in each absolute magnitude bin is shown as a solid
symbol. Solid errorbars mark the 25th to 75th percentile ranges of the distributions.
Dotted errorbars mark the 5th to 95th percentile ranges. As can be seen, the median
mass-to-light ratio first increases with luminosity and then appears to flatten. 
The scatter in $M/L(\rm{galaxy})$
at fixed luminosity decreases at longer wavelengths: it is more than  a factor of two larger
in the $g$-band than in the $z$-band. 

In order to make some comparison to related work,
we have plotted as a thick solid line the mean relation between $(M/L)_{r}(\rm{dynamical})$ and $r$-band magnitude
derived by Bernardi et al (2002) for a sample of early-type galaxies drawn from
the SDSS. In this analysis $M/L \propto R_0\sigma^2/L$, where $R_0$ is the effective
radius of the galaxy and $\sigma$ is the line-of-sight velocity dispersion of the stars 
measured in a 3 arcsecond aperture.  It is interesting that  the relation derived by
Bernardi et al agrees well with  our $M/L(\rm{galaxy})$ estimates for less luminous galaxies, but lies above
our estimates for brighter galaxies. This may be an indication that more luminous galaxies
contain more dark matter than less luminous galaxies. 
We caution, however, that a more careful analysis is required because the   
Bernardi et al sample is carefully selected to contain only galaxies with pure
early-type spectra, whereas our sample includes all galaxies. Note that if 
we had assumed a Salpeter  rather than a Kroupa (2001) IMF, our stellar mass estimates 
would have been around  a factor of two larger. For lower mass ellipticals,
they would then be clear
contradiction with the dynamically-derived mass estimates.

The distribution of formal errors on our stellar mass estimates is shown in Fig.14. 
We plot the distribution of the  95\% confidence range in $\log M_*$, i.e.
the range of values of $\log M_*$ obtained when the 2.5\% tails at
each end of the probability distribution of $\log M_*$ are excluded.
For a Gaussian distribution of  errors, this corresponds to four times the standard error  
in the mass estimate. As can be seen,  the 95\% confidence interval is typically
around a factor of $\sim 2$ in mass.
The errors are smaller
for older galaxies with larger 4000 \AA \hspace{0.1cm} breaks than for younger galaxies
with smaller break strengths.

\begin{figure}
\centerline{
\epsfxsize=15cm \epsfbox{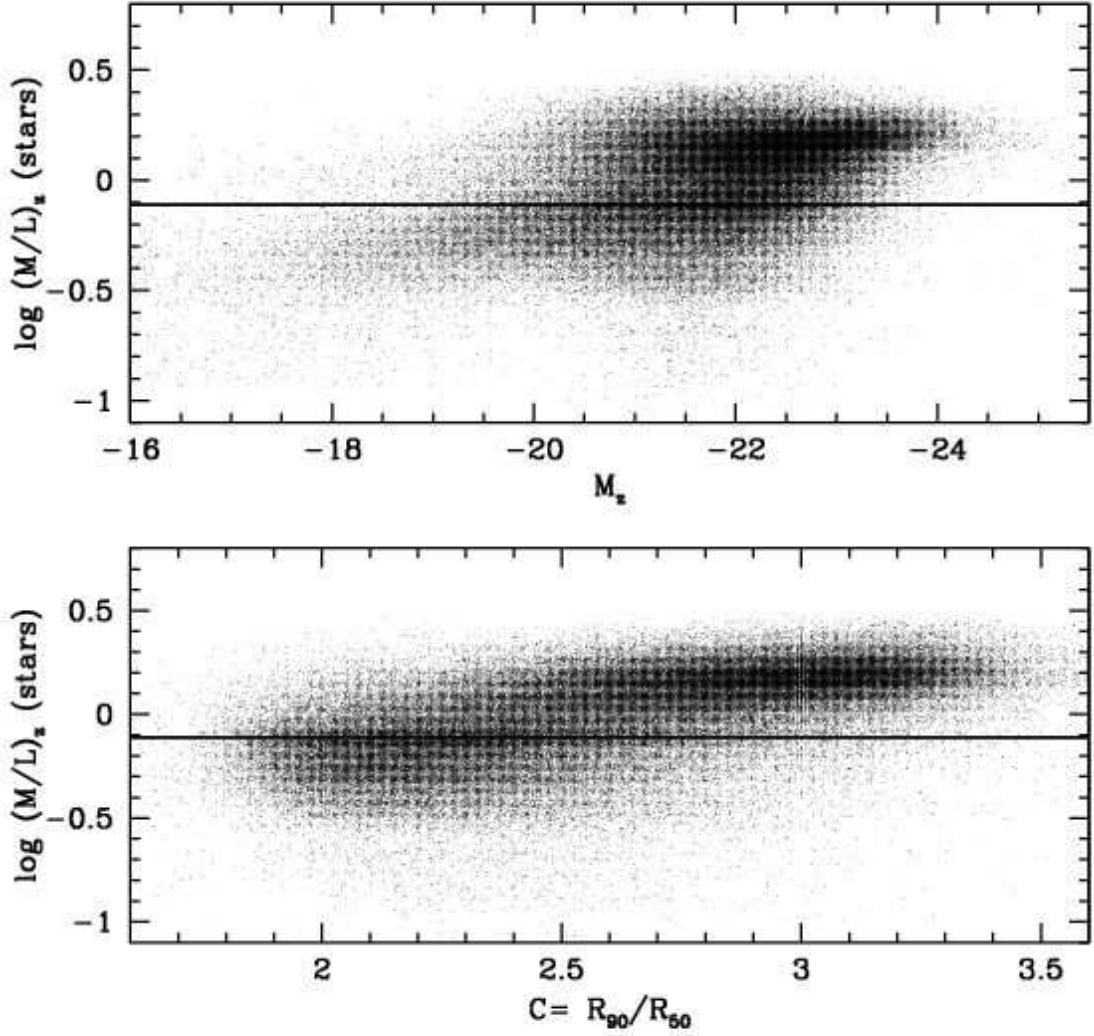}
}
\caption{\label{fig13}
\small
{\bf Top:} Our  estimate of $z$-band mass-to-light ratios of the stellar populations
of the galaxies in our sample is
 is plotted as a function of K-corrected  $z$-band absolute magnitude for all
galaxies in the sample.
{\bf Bottom:}  The mass-to-light ratio  is plotted as a function of $r$-band concentration parameter for all
galaxies in the sample.
The  line indicates the value of  $(M/L)_z(\rm{stars})$ for an unextincted galaxy 
that has been forming stars
at a constant rate for a Hubble time.}
\end {figure}
\normalsize

\begin{figure}
\centerline{
\epsfxsize=16cm \epsfbox{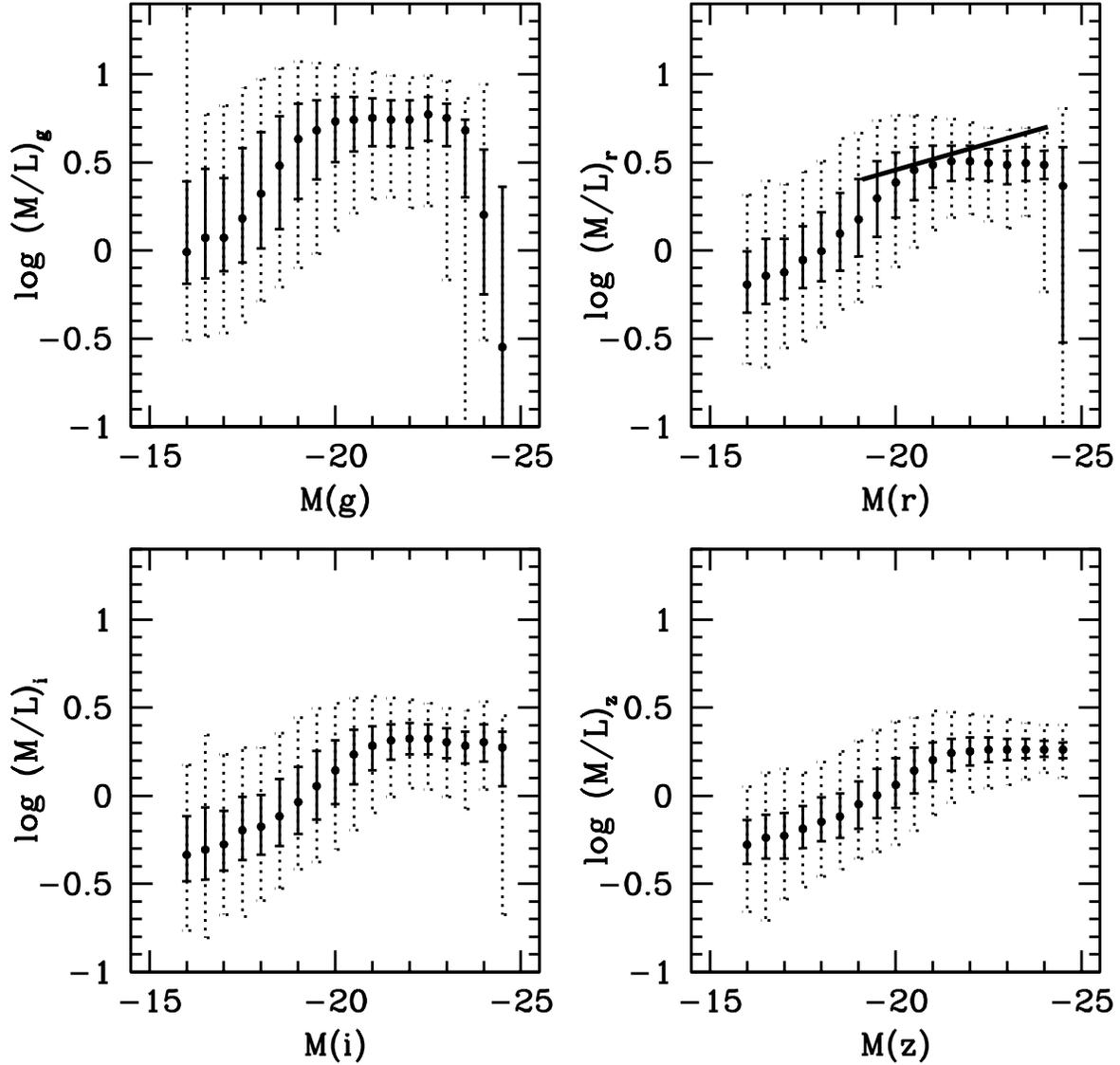}
}
\caption{\label{fig15}
\small
The total stellar mass of a galaxy divided by its observed luminosity (K-corrected to z=0.1) 
 is plotted as a function of absolute magnitude
in 4 SDSS bands. The solid symbols indicate the median mass-to-light ratio at a given
magnitude, the solid errorbars indicate the 25th to 75th percentile ranges of the
distribution and the dotted errorbars indicate the 5th-95th percentile ranges.
Mass-to-light ratios are plotted in solar units where $M_{\odot g}= 5.45$,
$M_{\odot r} = 4.76$, $M_{\odot i}= 4.58$ and $M_{\odot z}=4.51$ at $z=0.1$. (Blanton et al 2002).
The thick solid line is the mean relation between dynamical  
$M/L$ (estimated from the stellar velocity dispersion)
and $r$-band magnitude  for a sample
of early type galaxies from Bernardi et al (2002).} 
\end {figure}
\normalsize

\begin{figure}
\centerline{
\epsfxsize=9cm \epsfbox{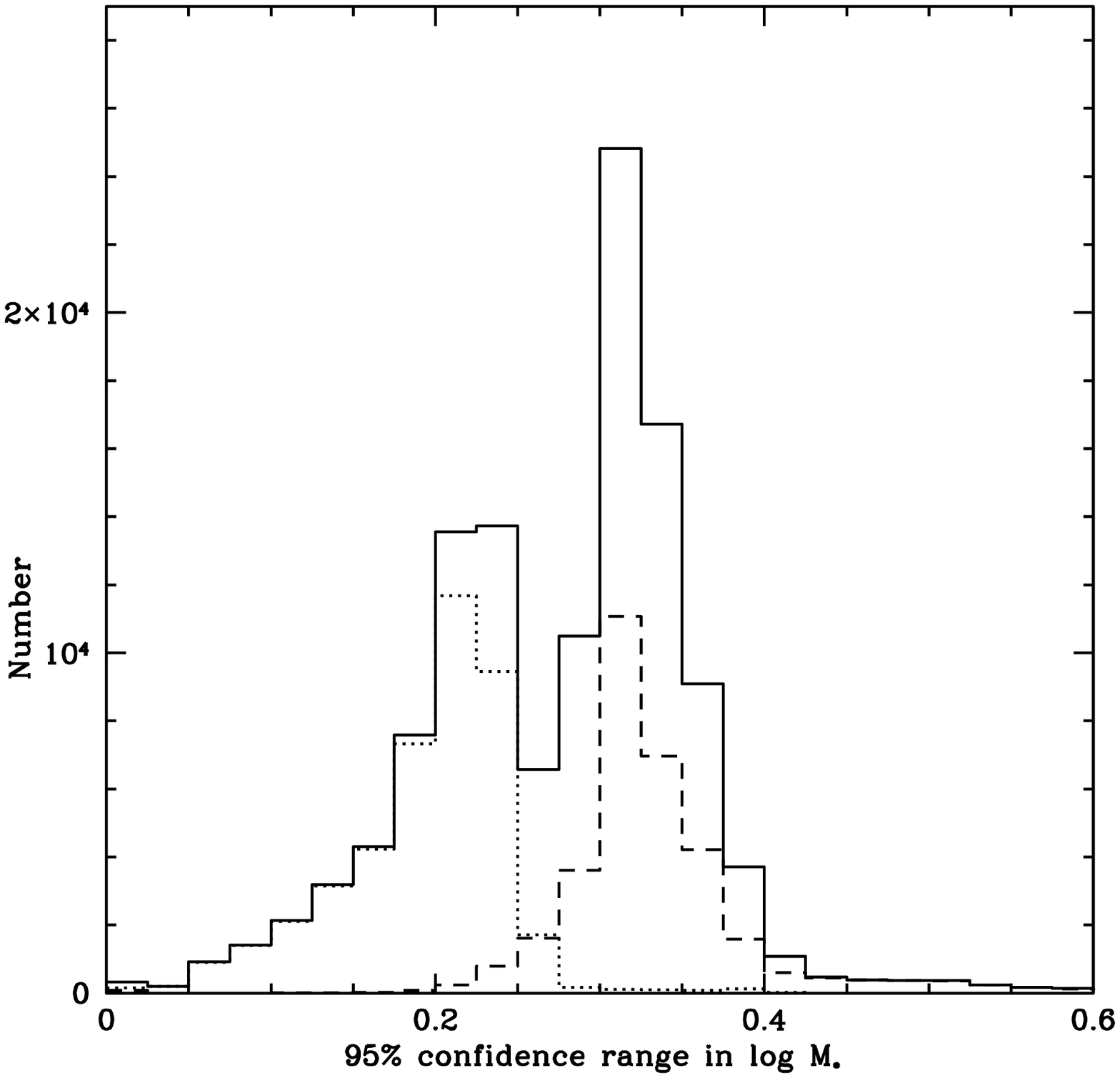}
}
\caption{\label{fig14}
\small
The distribution of errors in our stellar mass estimates. The quantity plotted is the
full length of the 95\% confidence interval in log M$_*$. The solid histogram
shows the distributions for all galaxies in the sample. The dotted histogram
is for the subset of galaxies with D$_n$(4000)$>$ 1.8 and the dashed
histogram is for galaxies with D$_n$(4000)$<$ 1.4.}
\end {figure}
\normalsize

\section {Sources of Systematic Error}

Fig. 14 indicates that the formal errors on our stellar mass estimates are small.
There are, however, quite a few sources of possible systematic error. In this
section, we explore some of these in more detail.

\subsection {Aperture effects}

The D$_n$(4000) and $H\delta_A$ indices measured from the SDSS spectra reflect the properties 
of the starlight that found its way down a 3 arcsecond diameter fibre that has been
positioned as close as possible to the centre of the galaxy. One may thus wonder to what
extent our estimates of $(M/L)_z$ may be biased because the index measurements are not
accurately reflecting the stellar content of the whole galaxy. In particular, aperture
bias could well be a serious problem for spiral galaxies, where the index measurements
may be appropriate for the central bulge component, but not for the outer disk where
most of the star formation is taking place.

It is possible to address the effect of aperture bias in a statistical way by comparing
the observed stellar indices and the derived $M/L$ values for similar galaxies 
viewed at different distances from us. If aperture bias is important, one should 
see a trend in these quantities with distance. 

This is illustrated in Figs. 16 and 17. We  plot the index  D$_n$(4000) and
the $z$-band mass-to-light ratio  $(M/L)_z$ as a
function of   `normalized' distance  $ z/z_{max}$, 
where $z_{max}$ is the redshift at which the galaxy 
drops out of the survey. 
Normalizing
by $z_{max}$  takes care of any selection effects that arise when one divides up the        
sample in different ways. Note that as expected,
galaxies in the survey are distributed uniformly
in the quantity $V/V_{max}$, so there are many more objects 
in the bins with  $z/z_{max}$ values close to 1.

Fig. 16 shows the variation in the 4000 \AA \hspace{0.1 cm} break
as a function of normalized distance. We see that both the size
and the sense of the effect  depend
strongly on the absolute magnitude of the galaxy.  
Galaxies with luminosities $\sim L_*$  ($M_*(z) =-22.3$) exhibit the strongest trend
with $z/z_{max}$. 
More distant galaxies
have younger stellar populations than nearby objects, as 
expected if the spectra are preferentially sampling the bulges of nearby
galaxies.                         

Fig. 17 shows how the dust attenuation in the z-band varies as a function of distance.
Our results indicate that $A_z$ increases towards the                             
centres of  low luminosity galaxies, but in galaxies with
luminosities $\sim L_*$, the attenuation increases towards the outer regions.
In the most luminous galaxies in our sample, the attenuation does not appear
to vary strongly as a function of radius. The simplest interpretation
of these trends is that our low-luminosity sample is dominated by disk
galaxies, which are known to have  have significant metallicity gradients (e.g.
Zaritsky, Kennicutt \& Huchra 1994). If dust content  correlates with
metallicity, one might expect the light from stars in the  inner parts of these galaxies to
be more attenuated. Galaxies with luminosities
$\sim L_*$ are composite systems consisting of both a bulge and a disk; starlight from the bulge is
expected to be less attenuated than that from the disk. The highest luminosity galaxies are primarily
ellipticals, which contain very little dust and have no appreciable gradients in $A_z$. 

Finally, Fig. 18 shows the trend in our estimates of $(M/L)_z$ with  distance.  
As can be seen, the effect of aperture on
the mass-to-light ratio is relatively small. For $L_*$ galaxies,
our estimate of the the median $(M/L)_z$ decreases by 0.12 dex from one edge of the survey to
the other. For brighter and fainter galaxies, the effect is even smaller.                                          
Unfortunately it is not possible use the relations shown in Figs. 16,  17 and 18
to estimate the true global  mass-to-light ratio of a given type of galaxy.
Even at the outer limit of the survey, the median fraction of the total galaxy light
that enters the fiber is around 50\%. A sample of galaxies with large aperture spectra is
required for this purpose.
Nevertheless, our plots make it clear  
that the variations in  $M/L$ as a function of luminosity and concentration 
are substantially larger than any biases induced by aperture effects.

\begin{figure}
\centerline{
\epsfxsize=7 cm \epsfbox{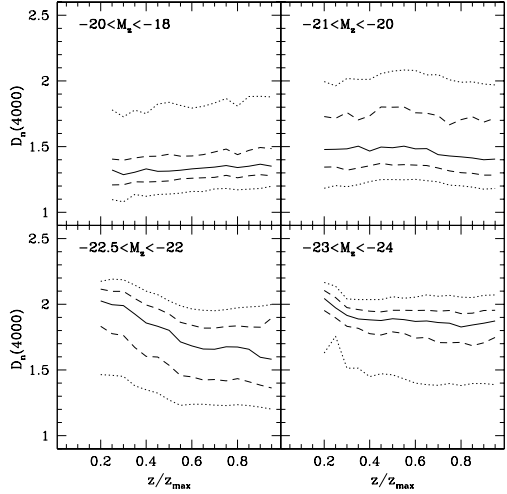}
}
\caption{\label{fig16}
\small
 The D$_n$(4000) index is plotted as a function of $z/z_{max}$ for galaxies in different ranges
of $z$-band absolute magnitude. The solid line indicates the median of the distribution as
a function of $z/z_{max}$. The dashed lines indicate the 25th and 75th percentiles. The dotted
lines indicate the 5th and 95th percentiles.}
\end {figure}
\normalsize

\begin{figure}
\centerline{
\epsfxsize=7 cm \epsfbox{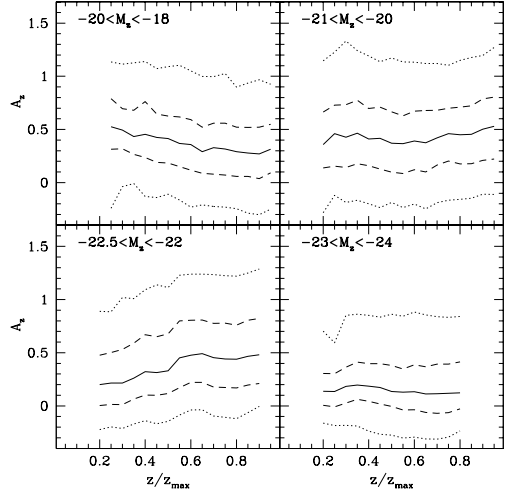}
}
\caption{\label{fig17}
\small
The dust attenuation in the $z$-band $A_z$  is plotted 
as a function of $z/z_{max}$ for galaxies in different ranges
of $z$-band absolute magnitude.}
\end {figure}
\normalsize

\begin{figure}
\centerline{
\epsfxsize=7 cm \epsfbox{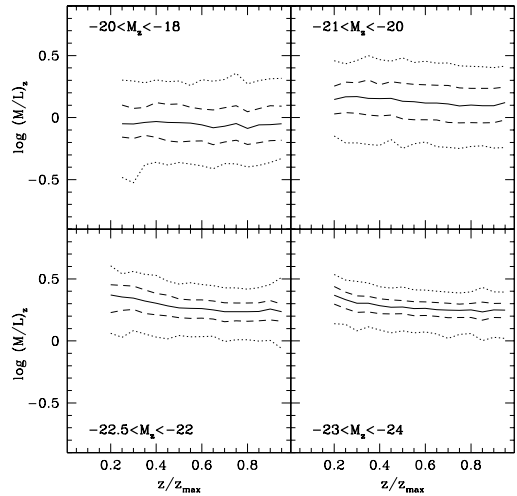}
}
\caption{\label{fig18}
\small
The stellar mass-to-light ratio in the $z$-band  is plotted 
as a function of $z/z_{max}$ for galaxies in different ranges
of $z$-band absolute magnitude.}
\end {figure}
\normalsize

\subsection {The choice of prior}

It is important to understand how our choice of priors may bias our estimates of
parameters such as mass-to-light ratios or burst mass fractions.
The assumed prior will have little influence if error bars are small and
the problem is well constrained. Because the errors on the H$\delta_A$ index
are of the same order as the total spread of values in the model grid,
parameters such as $F_{burst}$, which depend strongly on the value of
the H$\delta_A$ index, will be influenced by the mix of  star formation
histories contained in our model library.

This is illustrated in Fig. 19 where we compare, for two different choices of
prior,  our best estimates of  mass-to-light
ratio $(M/L)_z$,  attenuation $A_z$ and burst mass fraction $F_{burst}$.
P1 is our ``standard'' prior, described in detail in section 4.
P2 is a modified prior where  we have reduced by a factor of 10 the probability of
all models with bursts in the last 2 Gyr.
In the first three panels, we plot the median of the likelihood distribution obtained
using P2 versus the median obtained using P1. In the $F_{burst}$ panel, we only plot
those galaxies that have  $F_{burst}(50\%)>0$ for either P1 or P2.
As can be seen, our estimates of $(M/L)_z$ and $A_z$ are  insensitive to the
choice of prior. On the other hand, $F_{burst}$(median) does depend strongly on the
fraction of galaxies in the model grid that have experienced recent starbursts.

In the fourth panel, we compare the median of the likelihood distributions of
$F_{burst}$ for the {\em subset} of galaxies with $F_{burst}(2.5\%)>0$ for either prior, i.e.
we restrict the sample to galaxies that are very strongly constrained to have
experienced a recent starburst. The lower right panel of Fig. 19 demonstrates
that for these galaxies, $F_{burst}(50\%)$ 
is insensitive to to the choice of prior.

This demonstrates that it is valid to use the criterion $F_{burst}(2.5\%)>0$ to select  
subsamples of starburst and post-starburst galaxies from our full data set.
$F_{burst}(50\%)$ then provides the best estimate of the fraction of the total
mass in stars formed in the burst mode over the past 2 Gyr.

\begin{figure}
\centerline{
\epsfxsize=14 cm \epsfbox{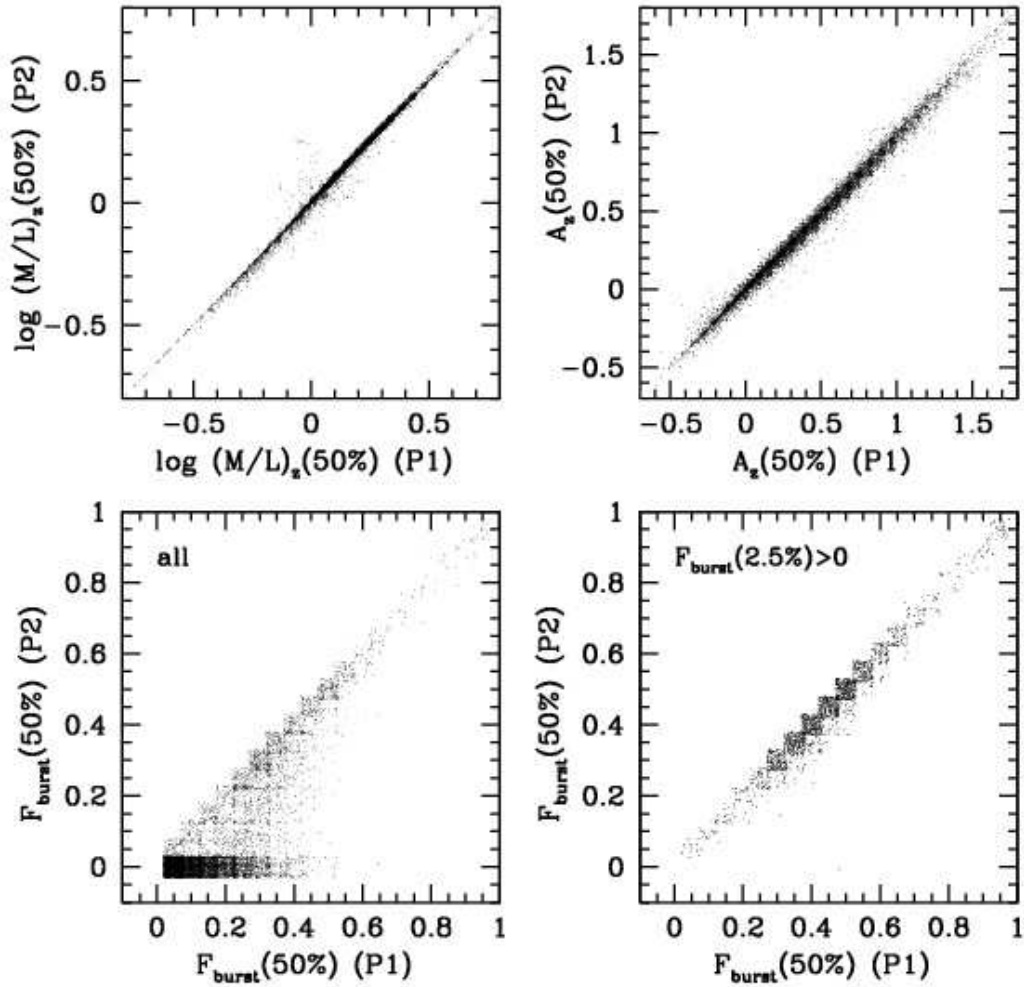}
}
\caption{\label{fig19}
\small
The derived values of the mass-to-light
ratio $(M/L)_z$, the attenuation $A_z$, the median of the likelihood distribution of
$F_{burst}$, and the lower 2.5 percentile value of this distribution are compared
for two different choices of prior.}
\end {figure}
\normalsize

\subsection {Other sources of systematic error}
\begin {enumerate}
\item {\em Stellar population models} 
Fig. 2 indicates that the  differences
in the predicted  values of D$_n$(4000) and H$\delta_A$ at given stellar age
for three different input stellar libraries are comparable to the
measurement errors for those indices. Unfortunately, we are only able to carry out
this test at solar metallicity, so we are unable to carry out a fully rigorous analysis
of the systematic uncertainty arising from our calibration of these indices.
\item {\em IMF.} All our derived parameters are tied to a specific choice of IMF. 
Changing the IMF would scale the stellar mass estimates by a
fixed factor. For example, changing from a Kroupa (2001) to a Salpeter IMF with a
cutoff at 0.1 $M_{\odot}$ would
result in a factor of 2 increase in the stellar mass.
\item {\em Calibration errors} Our dust corrections rely on accurately calibrated
spectral magnitudes. Any systematic offsets in the spectrophotometric
calibrations will result in systematic errors in these corrections. A full assessment
of the accuracy of the wavelength and flux calibration of SDSS spectra will be given in
Tremonti et al (2002).
\end {enumerate}

\section {Comparison with Colour-Based Methods}

Most previous attempts to convert from absolute magnitude
to stellar mass have used colours to constrain the  star formation histories
of galaxies (e.g. Brinchmann \& Ellis 2000; Cole et al 2001; Bell \& de Jong (2001)).
Bell \& de Jong (2001) argue that there should  be a tight relation         
between the optical colours of spiral galaxies and their stellar mass-to-light
ratios. This relation ought to be insensitive to the effects of dust-reddening.
Although dust causes colours to become redder, it also makes the galaxy
fainter and for a standard dust attenuation curve, the two effects compensate.
Bell \& de Jong (2001) show that recent bursts of star formation introduce
scatter into this relation, but claim that this should not be important for
most spirals in Tully-Fisher samples.

It is interesting to see whether the stellar masses derived using our method 
correlate with the observed optical colours of galaxies.  
In Fig. 20 we plot the stellar mass-to-light ratio in the $g$-band as a function
of the Petrosian $g-r$ colours (K-corrected to $z=0.1$) for galaxies in 4 different absolute
magnitude ranges. We find that there is a tight
relation between colour and stellar mass-to-light ratio. The scatter in log $(M/L)_g$ 
at given $g-r$ colour is roughly constant at all magnitudes and has an r.m.s. value 
of $\sim 0.3$ dex. However, the zero-point of the relation shifts systematically
by about 0.2 dex from the faintest to the brightest galaxies in our sample.   
At least part of this effect may be due to the fact that brighter galaxies have stronger
colour gradients (Fig. 16) and our mass-to-light will consequently be biased to
higher values than those measured from integrated colours.
Overall, the agreement is very encouraging, as it suggests that SDSS colours may be used to
derive a reasonably accurate estimate of the stellar masses of nearby galaxies when
spectroscopic information is missing. Note that the $r-i$ colour would {\em not} work
well because of its sensitivity to emission lines at redshifts $z \sim 0.1$.

\begin{figure}
\centerline{
\epsfxsize=16cm \epsfbox{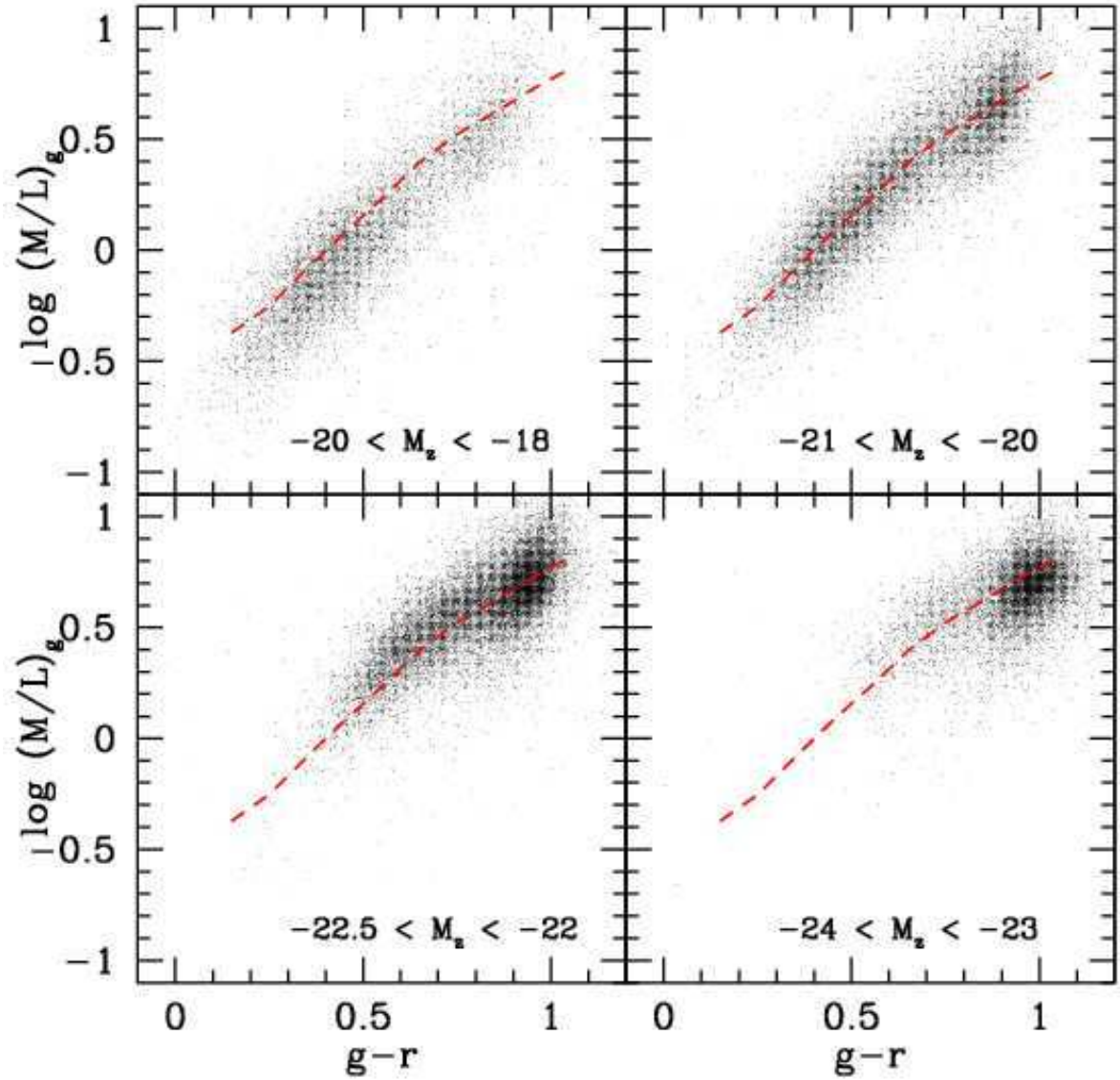}
}
\caption{\label{fig20}
\small
The $g$-band mass -to-light ratio is plotted as a function of the $g-r$
colour (K-corrected to z=0.1) for galaxies in 4 different 
bins of $z$-band absolute magnitude. The dashed
red line shows the mean relation evaluated for galaxies with
$-21 < M(z) <-20$. }
\end {figure}
\normalsize

\section {An Inventory of the Stellar Mass in the Universe}

In this section, we compute how galaxies of different types contribute to 
the total stellar mass budget of the local Universe.
Galaxies of different luminosities can be seen to different distances
before dropping out due to the selection limits of the survey.
The volume $V_{max}$  within which a galaxy can be seen and will be included in the sample
goes as the distance limit cubed, which results in galaxy
samples being dominated by intrinsically bright galaxies.
In this paper we use the simplest available  method for correcting for
selection effects, the $V_{max}$ correction method (Schmidt 1968).
Each galaxy is given a weight equal to the inverse of its maximum
visibility volume determined from the apparent magnitude limit of the survey.
In order to obtain an estimate of the true number density of galaxies in the Universe,
it is necessary to account for 
galaxies that are missed due to fibre collisions or 
spectroscopic failures. This affects around 10-15\% of the galaxies brighter
than the spectroscopic limit of the survey (Blanton et al 2001).
Here, we focus on the {\em relative} contribution of different kinds of galaxies
to the total stellar mass, so weighting by $1/V_{max}$ ought to be sufficient.
As a check, we have computed the $z$-band luminosity function for the galaxies
in our sample and find that the Schechter function fit given in Blanton et al (2001)
provides an excellent match to the shape of the function we derive.

A compendium of results is presented in Fig. 21. We plot the fraction of the
total stellar mass contained in galaxies as a function of their stellar mass, D$_n$(4000),
$g-r$ colour, $F_{burst}$, size, concentration index and surface mass density.  
Note that we have adopted a cosmology with $\Omega=0.3$, $\Lambda=0.7$
and H$_0$= 70 km s$^{-1}$ Mpc$^{-1}$ in our calculations.
From these plots it is possible to read off the characteristic properties
of the galaxies that contain the bulk of the stellar mass in the Universe at
the present day. Note that because we have  122,808 galaxies in the total
sample, we are able to derive extremely accurate functional forms for these
distributions. The bins around the peaks of the distributions shown in
Fig. 21 each contain 5000 or more galaxies. Even in the wings, most bins 
are still sampled by a few hundred objects. In Table 1, we list the median,
1\%, 5\%, 25\%, 75\%, 95\% and 99\%  percentile values  
for each of the distributions shown in Fig. 21 
(with the exception of $F_{burst}$).

Our main conclusions are the following:
\begin {itemize}
\item  The characteristic  mass $M_{char}$ of galaxies at the present day, defined  
as the peak of the distribution function shown in the top
left panel of Fig. 21,
is $6 \times 10^{10} M_{\odot}$. This agrees reasonably well with the results 
of  Cole et al. (2001). These authors transform the near-infrared luminosity
function derived from 2dF/2MASS data to a stellar mass function using
a colour-based technique. Their Schechter-function parametrization  yields
$M_{char} = 5.6 \times 10^{10} M_{\odot}$ for H$_0$= 70 km s$^{-1}$ Mpc$^{-1}$
and a Kennicutt (1983) IMF, which is fairly close to the Kroupa (2001) IMF
that we have assumed.
 
We find that only 20\% of the total stellar mass is contained
in galaxies less massive than $10^{10} M_{\odot}$. An even
smaller fraction ($\sim$1\%) of the total
mass is contained in galaxies less massive than $10^{9} M_{\odot}$. 
The total stellar mass budget of the Universe is thus heavily weighted towards galaxies
that are within a factor of 10 in mass of the Milky Way.  
\item The D$_n$(4000) distribution of the stellar mass is strongly bimodal.
The first peak is centred at D$_n$(4000)$\sim 1.3$.
Galaxies with break strengths of this value
have $r$-band weighted mean
stellar ages of $\sim 1-3$ Gyr and mass-weighted
mean ages a factor of $\sim 2$ larger. Almost all these galaxies also
have emission lines and are thus forming stars at the present day.
The second peak is centred at D$_n$(4000)$\sim 1.85$, a value  typical
of old elliptical galaxies with mean stellar ages $\sim 10$ Gyr.
\item The $g-r$ colour distribution exhibits a strong red peak, but 
the blue peak is much less pronounced. This is probably because
colours depend both on stellar age and on dust attenuation, 
whereas D$_n$(4000) is not affected by dust. Note that Strateva et al (2001) have also
discussed the  bimodality in the colour distributions of SDSS galaxies.     
\item The vast majority ($> 90 \%$) of the stellar mass is in galaxies  that have  not
formed more than 5\% of the stars in a burst in the past 2 Gyr. 
\item The distribution of stellar mass as a function of 
galaxy  half-light radius in the $r$-band ($R_{50}$)              
is peaked at $\sim 3$ kpc.
For the radius containing  90\% of the light ($R_{90}$) it is peaked at $\sim 8$ kpc.
More than 90\% of the total stellar mass in the Universe resides in galaxies with
$R_{50}$ and $R_{90}$ that are within a factor of three of these values.
\item The distribution of stellar mass as a function of concentration index is
broad, with no pronounced peak at any particular value.
If we adopt $C=2.6$ as the demarcation between early and late-type galaxies,
then 50\% of the total stellar mass is contained in the early-types.
If we adopt $C=3$, then only 10\% of the mass is contained in such
systems.
\item We define the surface mass density $\mu_*$ as $0.5M_*/(\pi z_{50}^2)$,
where $z_{50}$ is the Petrosian half-light radius in the $z$-band. 
Most of the stellar mass in the Universe resides in galaxies with
$\mu_*$ within a factor of 2 of  $10^9 M_{\odot}$ kpc$^{-2}$.
\end {itemize}

\vspace {2.5 cm}
\normalsize

{\bf Table 1:} Percentiles of the distribution of the fraction of the 
total stellar mass contained in galaxies of different types shown in Fig. 21                      
\vspace {0.3cm}

\begin {tabular} {lccccccc}
Parameter  & Median & 1\% & 5\% & 25\% & 75\% & 95\% & 99\% \\                                   
log $M_*$ (M$_{\odot})$  & 10.493 & 9.249&  9.542 & 10.108 & 10.799 & 11.139 & 11.342 \\
D$_n$(4000) & 1.609 & 1.105&  1.120 & 1.350 & 1.838& 2.012 & 2.119 \\
$g-r$ (z=0.1) & 0.816 & 0.238&  0.380& 0.634 & 0.906& 1.095& 1.411 \\
$R_{50}$ (kpc) & 3.225 & 0.249 & 0.740 & 2.042 & 4.636 & 7.190 & 9.652 \\
$R_{90}$ (kpc) & 8.748 & 0.784 & 2.075 & 5.417 & 12.177 & 19.386 & 27.014 \\
$C= R_{90}/R_{50}$ & 2.493 & 1.767&  1.896 & 2.201 & 2.787& 3.106& 3.284 \\
log $\mu_*$ (M$_{\odot}$ kpc$^{-2}$) & 8.745 & 7.327& 7.810 & 8.453 & 8.958 & 9.242& 9.484 \\
\end {tabular}

\vspace {1.5 cm}

\begin{figure}
\centerline{
\epsfxsize=15cm \epsfbox{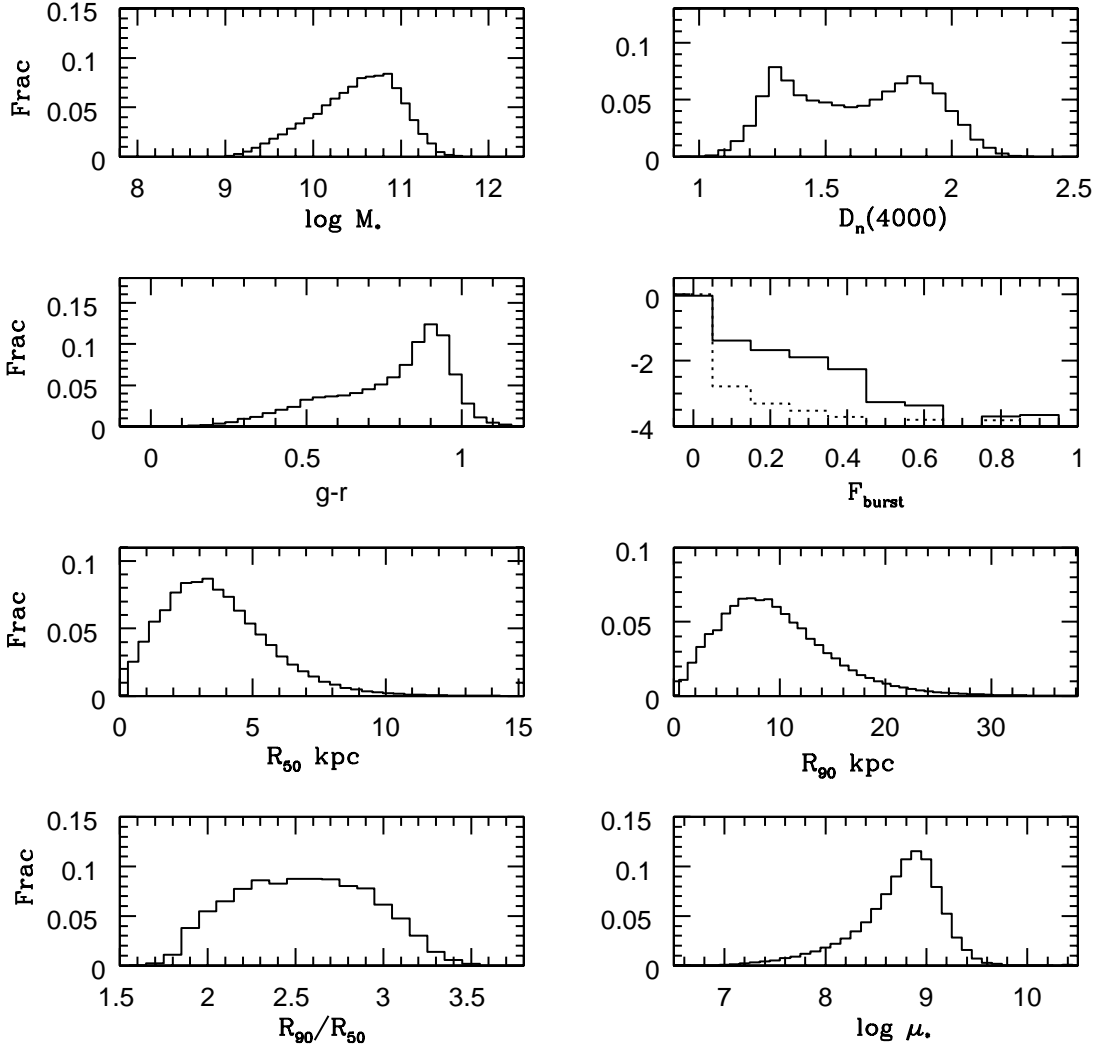}
}
\caption{\label{fig21}
\small
The fraction of the total stellar mass in the Universe contained in galaxies
as a function of 1) log stellar mass, 2) D$_n$(4000), 3) $g-r$ colour
K-corrected to z=0.1, 4) $F_{burst}$ (median) {\em solid} and $F_{burst}$(2.5\%) {\em dotted}
 , 5) Petrosian half-light radius in the
$r$-band, 6) Petrosian 90\% radius on the $r$-band, 7) concentration
index ($R_{90}/R_{50}$), 8) log surface mass density. The fraction is shown linearly in all
plots except that for $F_{burst}$ where the logarithm is given.}
\end {figure}
\normalsize

\section {Summary and Discussion}

We have developed a new method to constrain the past star formation histories of 
galaxies. It is based on two stellar absorption line indices, the 4000 \AA \hspace{0.1cm}
break strength D$_n$(4000) and the Balmer absorption line index H$\delta_A$.
Together these two indices allow us to place constraints on the mean age
of the stellar population of a galaxy and the fraction of its stellar mass
formed in recent bursts. 

We have generated a library of Monte Carlo realizations 
of different star formation histories, which includes bursting as well as
continuous models and a wide range of metallicities. 
We use the library to generate estimates and confidence intervals for
a variety of  parameters for a sample of 122,808 galaxies drawn from
the Sloan Digital Sky Survey, These include:
\begin {enumerate}
\item $F_{burst}$ the fraction of the stellar mass of the galaxy formed
 in bursts in the past two Gyr.
\item $A_{z}$ the attenuation of the rest-frame $z$-band light due to dust.
\item Stellar mass-to-light ratios in the $g$, $r$, $i$ and $z$ bands.
\item Stellar masses.
\end {enumerate}

Note that the analysis can be extended to include other parameters describing
the star formation history of a galaxy, but not all parameters are equally
well-constrained. For example, the luminosity-weighted or mass-weighted mean stellar 
ages have large errors, because of the rather strong dependence of the 4000 \AA 
\hspace{0.1cm} break on metallicity at ages of more than  1-2 Gyr (see Fig. 2).

In the first part of the paper, we have illustrated in detail how our 
methods can be applied to galaxies in the SDSS
and we have estimated the confidence with which we can constrain
basic parameters such as stellar mass. 
In the second part of the paper, we have presented
a number of astrophysically interesting applications of our methods.

We have shown that the attenuation of the $z$-band light due to dust depends
strongly on both the absolute magnitude and the mean stellar age of a galaxy.
We have also studied the distribution of stellar mass-to-light ratios of galaxies
as a function of absolute magnitude in the four SDSS pass bands. 
We have shown that the distribution of $M/L$ is strongly dependent on galaxy luminosity in
all photometric bands.
Almost all very luminous galaxies have high mass-to-light ratios. 
Faint galaxies have lower mass-to-light ratios, but also span a  wider range in $M/L$. 
We have also shown that the scatter in $M/L$ at a given luminosity is a factor
of $\sim 2$ smaller in the $z$-band than it is in the $g$-band.
Finally, we have computed how galaxies of different types contribute to the total
stellar mass budget of the Universe.

In the standard paradigm for structure formation in the Universe, galaxy
formation occurs hierarchically through the merging of small
protogalactic condensations to form more and more massive systems.
In this picture, the contribution of different kinds of galaxies to
the stellar mass budget is expected to evolve strongly with redshift.
The rate and form of this evolution  
depend not only on the values of cosmological parameters such as
$\Omega$ and $\Lambda$, but also on the physical processes that 
control the rate at which stars form in galaxies.
For example, if star formation rates
are enhanced during galaxy-galaxy mergers, the fraction of stars
formed in recent bursts should rise strongly with redshift, simply 
because merging rates and gas fractions were higher in the past (see for
example Kauffmann \& Haehnelt 2000). 

Although it is now clear that the integrated star formation rate density
increases strongly to higher redshifts (Madau et al 1996), it is still
not understood which galaxies undergo the strongest evolution or which 
physical processes cause galaxies to form stars more rapidly in the past.
In the next few years, there will be a number of new large redshift 
surveys of the faint galaxy population. These surveys will contain enough galaxies
to carry out an inventory of the stellar mass
at $z \sim 1$. When these results  are compared with the distributions derived from the
SDSS, it will be possible to draw  
quantitative conclusions about how
galaxies have evolved over the past two thirds of a Hubble time and to begin 
disentangling the effects of the different processes that may have influenced this
evolution. 

\vspace {1cm}

We thank the anonymous referee for detailed comments that helped improve our methodology.

S.C. thanks the Alexander von Humboldt Foundation, the Federal Ministry of Education
and Research, and the Programme for Investment in the Future (ZIP) of the German
Government for their support.

 The Sloan Digital Sky Survey (SDSS) is a
joint project of The University of
           Chicago, Fermilab, the Institute for
                              Advanced Study, the Japan Participation
                              Group, The Johns Hopkins University, 
                              Los Alamos National Laboratory, the
                              Max-Planck-Institute for Astronomy
                              (MPIA), the Max-Planck-Institute for
                              Astrophysics (MPA), New Mexico State
                              University, Princeton University, the
                              United States Naval Observatory, and the
                              University of Washington. Apache Point
                              Observatory, site of the SDSS
                              telescopes, is operated by the
                              Astrophysical Research Consortium (ARC).

                              Funding for the project has been
                              provided by the Alfred P. Sloan
                              Foundation, the SDSS member
                              institutions, the National Aeronautics
                              and Space Administration, the National
                              Science Foundation, the U.S. Department
                              of Energy, the Japanese Monbukagakusho,
                              and the Max Planck Society. The SDSS Web
                              site is http://www.sdss.org/.

\pagebreak
\normalsize
\Large
{\bf Appendix A}\\
\normalsize

\vspace {0.3 cm}
Here we describe the mathematical 
underpinning of the Bayesian
likelihood estimates for parameters such as $F_{burst}$, $A_{z}$ and
$M/L$  that we derive in this paper. 

Let us recall the basis of Bayesian statistics for this kind of problem.
An initial assumption is made that the data are randomly drawn from
a distribution which is a member of a model family characterised
by a parameter vector {\bf P}. The dimension of {\bf P} can, in principle, be
arbitrarily large. In particular, it can be much larger than the number 
of points N in the dataset to be fitted. The goal is then to use the data
to define a likelihood function on the space of all possible {\bf P}. This
function can be used to obtain a best estimate and confidence interval
for any model property Y({\bf P}).

 In Bayesian statistics one has to specify a prior distribution on
the space of all possible {\bf P}. This is a probability density distribution
$f_p$({\bf P}) which encodes knowledge about the relative likelihood of 
various {\bf P}
values in the absence of any data. For example, the physically accessible
range for each element of {\bf P} may be limited. Typically one takes a uniform
prior in parameters with a small dynamic range and a uniform prior in the
log of parameters with a large dynamic range.

The likelihood of a particular value of {\bf P} given a specific dataset ${\bf d}$ is
then written as a posterior probability density function (using Bayes' theorem)
as

\begin{equation} f({\bf P} \mid {\bf d}) d{\bf P} 
 = A f_p({\bf P}) Pr\{{\bf d} \mid {\bf P} \} d{\bf P} \end{equation}
where A is a constant which is adjusted so that $f$({\bf P} $\mid {\bf d})$ normalises
correctly to unity and $Pr\{ {\bf d} \mid {\bf P} \}$ is the probability of the observed
dataset on the hypothesis that the underlying distribution is described
by the particular parameter set {\bf P}.

The likelihood of the derived parameter Y({\bf P}) given the data is then
\begin {equation} f(Y \mid {\bf d}) dY  = \int_Y f({\bf P} \mid {\bf d}) d{\bf P} \end {equation}
where the integral extends over all {\bf P} for which Y lies in a specified
bin $\pm$dY/2. Note that there is no regularity requirement on the function
Y({\bf P}) other than piecewise continuity so it makes sense to define
a probability density. The most likely value of Y can then be taken as
the peak of this distribution; the most typical value as its median;
the 95\% (symmetric) confidence interval for scalar Y  can be defined
by excluding the 2.5\% tails at each end of the distribution.

 When applied to the kinds of problems presented in this
paper, the prior is  taken to be
the distribution of possible star formation histories
in the comparison library, which can be viewed as a Monte Carlo sampling
of $f_p({\bf P})$. 
The integral in the above expression for the
likelihood of Y is then trivially evaluated through binning the integrand
as a function of Y. The expression for $Pr\{{\bf d} \mid {\bf P} \}$ is also straightforward
for our case, since we have a measure of each element of {\bf d} and we
can assume the errors are normal with known correlation matrix C.
In this situation

\begin {equation} Pr\{{\bf d} \mid {\bf P} \}  \propto 
 \exp [ -({\bf d} - {\bf d}_p({\bf P})).{\rm C}^{-1}.({\bf d} - {\bf d}_p({\bf P}))/2] \end {equation}
where ${\bf d}_p({\bf P})$ is the data vector predicted by the model with parameters {\bf P}.
The argument of the exponential is then just minus one half of
$\chi^2$.

It is worth noting that this procedure makes no assumptions about the
shapes of the distributions $f_p({\bf P})$, $f({\bf P} \mid  {\bf d})$ 
or $f(Y \mid {\bf d})$. The first can
be assumed at will, and for small error bars and a well constrained problem
should have little effect on the answer. The other two are then derived
consistently. The important assumptions are that the model makes a
well defined and specific prediction for the value of the observable
in the absence of observational errors (in practice there will be
some degree of theoretical
uncertainty  and this could be included in C 
if it can be modelled
as Gaussian) and that the observed data {\bf d} have a known observational
error which can be assumed Gaussian with covariance matrix C.

\pagebreak 
\Large
\begin {center} {\bf References} \\
\end {center}
\normalsize
\parindent -7mm  
\parskip 3mm

Balogh, M.L., Morris, S.L., Yee, H.K.C., Carlberg, R.G., Ellingson, E., 1999, 
ApJ, 527, 54 

Bell, E.F., De Jong, R., 2001, ApJ, 550, 212

Bernardi, M., Sheth, R.K., Annis, J., Burles, S., Eisenstein, D.J., Finkbeiner, D.P.,
Hogg, D.W., Lupton, R.H.,  Schlegel, D.J., SubbaRao, M.  et al., 2002, AJ, 
submitted (astro-ph/0110344)

Blanton, M.R., Dalcanton, J., Eisenstein, D., Loveday, J., Strauss, M.A., SubbaRao, M.,
Weinberg, D.H., Andersen, J.E. et al., 2001, AJ, 121, 2358

Blanton, M.R., Brinkmann, J., Csabai, I., Doi, M., Eisenstein, D., Fukugita, M., Gunn, J.E.,
Hogg, D.W., Schlegel, D.J., 2002, AJ, submitted 

Blanton, M.R. et al., 2002, SDSS preprint

Brinchmann, J., Ellis, R.S., 2001, ApJ, 536, L77

Bruzual, A.G., 1983, ApJ, 273, 105

Bruzual, A.G., Charlot, S., 1993, ApJ, 405, 538

Calzetti, D., Kinney, A., Storchi-Bergmann, T., 1994, ApJ, 429, 582

Cayrel de Strobel, G, Hauck, B., Francois, P., Thevenin, F., Friel, E., Mermilliod, M., Borde, S., 1992, A\&AS,
95, 273

Charlot, S., Fall, S.M., 2000, ApJ, 539, 718

Cole, S., Norberg, P., Baugh, C.M., Frenk, C.S., Bland-Hawthorn, J., Bridges, T., 
Cannon, R., Colless, M. et al, 2001, MNRAS, 326, 255

Fukugita, M., Ichikawa, T., Gunn, J.E., Doi, M., Shimasaku, K., Schneider, D.P., 1996,
AJ, 111, 1748

Gorgas, J., Faber, S.M., Burstein, D., Gonzalez, J.J., Courteau, S., Proseer, C., 1993, ApJS, 86, 153

Gorgas, J., Cardiel, N., Pedraz, S., Gonzalez, J.J., 1999, A\&AS, 139, 29

Gunn, J.E., Carr, M., Rockosi, C., Sekiguchi, M., Berry, K., Elms, B., de Haas, E.,
Ivezic, Z. et al, 1998, AJ, 116, 3040 

Hogg, D.W., Finkbeiner, D.P., Schlegel, D.J., Gunn, J.E., 2001, AJ, 122, 2129

Jacoby, G.H., Hunter, D.A., Christian, C.A., 1984, ApJS, 56, 257

Kauffmann G., Haehnelt M., 2000, MNRAS, 311, 576

Kauffmann, G., Heckman, T.M., White, S.D.M., Charlot, S., Tremonti,C., Peng, E.W., Seibert, M.,
Brinkmann, J. et al, 2002, MNRAS, submitted (astro-ph/02005070) 

Kennicutt, R.C., 1983, ApJ, 272, 54

Kroupa, P., 2001, MNRAS, 322, 231

Le Borgne, J.-F., Bruzual, A.G., Pello, R., Lancon, A., Rocca-Volmerange, B., Sanahuja, B.,
Schaerer, D., Soubiran, C., Vilchez-Gomez, R., 2002, in preparation (http://webast.ast.obs-mip.fr/stelib)

Liu, M.C., Charlot, S., Graham, J.R., 2000, ApJ, 543, 644 

Lupton, R.H. et al, 2002, in preparation

Madau, P., Ferguson, H.C., Dickinson, M.E., Giavalisco, M., Steidel, C.C., 
Fruchter, A., 1996, MNRAS, 283, 1388 

McKay, T.A., Sheldon, E.S., Racusin, J., Fischer, P., Seljak, U., Stebbins, A.,
Johnston, D., Frieman, J.A. et al., 2002, ApJ, submitted (astro-ph/0108013)

Pickles, A.J., 1998, PASP, 110, 863

Schlegel, D.J., Finkbeiner, D.P., Davis., M., 1998, ApJ, 500, 525

Shimasaku, K., Fukugita, M., Doi, M., Hamabe, M., Ichikawa, T., Okamura, S., 
Sekiguchi, M., Yasuda, N. et al, 2001, AJ, 122, 1238

Schmidt, M., 1968, ApJ, 151, 393

Smith, J.A., Tucker, D.L., Kent, S., Richmond, M.W., Fukugita, M.,
Ichikawa, T., Ichikawa, S.I.,Jorgensen, A.M.  et al. 2002, AJ, in press

Stoughton, C., Lupton, R.H., Bernardi, M., Blanton, M.R., Burles, S., Castander, F.J.,
Connolly, A.J., Eisenstein, D.J. et al., 2002, AJ, 123, 485

Strateva, I., Ivezic, Z., Knapp, G.R., Narayanan, V.K., Strauss, M.A., Gunn, J.E., 
Lupton, R.H., Schlegel, D. et al, 2001, AJ, 122, 1861 

Strauss, M.A., Weinberg, D.H., Lupton, R.L., Narayanan, V.K., Annis, J., Bernardi, M.,
Blanton, M., Berles, M. et al., 2002, AJ, accepted (astro-ph/0206225)

Tremonti, C.A. et al., 2002, in preparation

Verheijen, M.A.W., 2001, ApJ, 563, 694

Wang, B., Heckman, T.M., 1996, ApJ, 457, 645

Worthey, G., Ottaviani, D.L, 1997, ApJS, 111, 377

York D.G., Adelman J., Anderson J.E., Anderson S.F., Annis J., Bahcall N.A., Bakken J.A.,
Barkhouser R. et al., 2000, AJ, 120, 1579

Zaritsky, D., Smith, R., Frenk, C.S., White, S.D.M., 1993, ApJ, 405, 464

Zaritsky, D., Kennicutt, R.C.. Huchra, J.P., 1994, ApJ, 420, 47

\end{document}